\documentstyle[eqsecnum,aps,multicol,pre]{revtex} 
\input epsf
\newcommand{\dir}{Figs}

\newcommand{\fig}[4]
{
     \noindent
     \unitlength=1mm
     \begin{picture}(#2,#3)
     \put(10,0){\leavevmode \epsfxsize=#2mm \epsffile{\dir/#1}}
     \end{picture}
   \noindent
#4
}
\begin{document} 

\newcommand{\CCphbulk}
{
\caption{ Schematic phase diagram of the symmetrical binary fluid
mixture in the density-temperature plane. Dashed line indicates
critical demixing transitions, full curve the first order 
liquid-vapour coexistence envelope.}
\label{fig:phbulk}
}

\newcommand{\CCprofschem}
{
\caption{
Schematic sketch of density and order parameter profiles in the limit 
$\gamma/{m_{+}^*}^2 \to \infty$. See text for more explanation.}
\label{fig:profschem}
}

\newcommand{\CClg}
 {
\caption{ Length scales $\lambda_1$ and $\lambda_2$ vs. $\gamma/{m_+^*}^2$
at $\kappa=0.5$. Thin line shows $\lambda_1/2$ for comparison.}
\label{fig:lg}
}

\newcommand{\CCdg}
{
\caption{ Scaling variable $\delta$ vs. $\gamma/{m_+^*}^2$ at $\kappa = 0.5$}
\label{fig:dg}
}

\newcommand{\CCtg}
{
\caption{ Coefficients $\tau_i$ of interfacial potential vs. 
$\gamma/{m_+^*}^2$ for $\kappa=0.5, \Phi_0 = 1.5$, and surface coupling
$C_m = 0.$ (thick lines) and $C_m = 0.2$ (thin lines).}
\label{fig:tg}
}

%\newcommand{\CCvla}
%{
%\caption{ Effective interface potential for different surface densities 
%$\rho_0$ at $\gamma/{m_+^*}^2=7.5, C_m=0.$}
%\label{fig:vl1}
%}

%\newcommand{\CCvlb}
%{
%\caption{ Effective interface potential at the wetting transition for 
%different $\gamma/{m_+^*}^2$.}
%\label{fig:vl2}
%}

%\newcommand{\CCvlc}
%{
%\caption{ Effective interface potential at the wetting transition for 
%different $C_m$ ($\gamma/{m_+^*}^2 = 7.5$).}
%\label{fig:vl3}
%}

\newcommand{\CCphweta}
{
\caption{ 
Phase diagram in the $\mu-\phi_0$ plane. Parameters
are $\kappa=0.5, \theta=0.1$ and $\gamma =1., C_m = 0.2$.
Solid line indicates first order transition, dashed line second order 
transition.}
\label{fig:phwet1}
}

\newcommand{\CCprofiles}
{
\caption{ Density and order parameter profiles for the coexisting
mixed and demixed film at the point in the $\mu-\phi_0$ plane
indicated by the arrows in Fig. 6.}
\label{fig:profiles}
}

\newcommand{\CCphwetb}
{
\caption{ Phase diagrams in the $\mu-\phi_0$ plane for different
$\gamma$ with $\theta=0.1$, surface coupling $C_m=0.$, and parameters 
as in Fig. 6 otherwise. Solid lines indicate first order transition, 
dashed lines second order transitions.}
\label{fig:phwet2}
}

\newcommand{\CCphwetc}
{
\caption{ Phase diagrams in the $\mu-\phi_0$ plane for different
$\theta$ with $\gamma=1.$, surface coupling $C_m=0.$ and parameters as 
in Fig. 6 otherwise. Solid lines indicate first order transition, dashed 
lines second order transitions.}
\label{fig:phwet3}
}

\newcommand{\CCmuT} 
{ 
\caption{ The phase diagram in the $\mu$-$T$ plane
of the symmetrical binary Lennard-Jones fluid model described in the
text.  Also shown is the location of the critical end point and the
isotherm along which the wetting properties were studied} \label{fig:muT}
}

\newcommand{\CCewA}
{
\caption{Density profiles for $\epsilon_w=1.0$. Data are shown
for $7$ values of $\mu-\mu_{cx}$ in the range $[0,-1.5]$ } \label{fig:ew1.0}
}

\newcommand{\CCewB}
{
\caption{ Density profiles for $\epsilon_w=1.7$. Data are shown
for $8$ values of $\mu-\mu_{cx}$ in the range $[0,-1.525]$ } \label{fig:ew1.7}
}

\newcommand{\CCewC}
{
\caption{ Density profiles for $\epsilon_w=1.75$. Data are shown
for $6$ values of $\mu-\mu_{cx}$ in the range $[-0.025,-1.5]$ } \label{fig:ew1.75}
}

\newcommand{\CCewD}
{
\caption{ Density profiles for $\epsilon_w=2.0$. Data are shown
for $8$ values of $\mu-\mu_{cx}$ in the range $[-0.025,-1.6]$ } \label{fig:ew2}
}

\newcommand{\CCewE}
{
\caption{{\bf (a)} Density profiles for $\epsilon_w=3.0$. Data are shown
for $6$ values of $\mu-\mu_{cx}$ in the range $[-0.025,-1.55]$ } \label{fig:ew3}
{\bf (b)} The corresponding order parameter profiles n(z). }

\newcommand{\CCewF}
{
\caption{ Density profiles for $\epsilon_w=4.0$. Data are shown
for $8$ values of $\mu-\mu_{cx}$ in the range $[-0.025,-1.6]$ } \label{fig:ew4}
}

\newcommand{\CCwetschem}
{
\caption{ Some possible schematic wetting phase diagrams in the 
temperature-density plane. 
(a) Weakly attractive wall: Critical end point $T_{cep}$ of the 
$\lambda$-line is a critical wetting point, below which the wall 
is not wetted by the liquid. 
(b) Intermediate attraction: Demixing induced first order wetting transition 
at $T < T_{cep}$ with prewetting line which evolves into a second
order demixing line.
(c) Strong attraction: Complete wetting at coexistence everywhere,
but detached prewetting line or continuous demixing transitions 
off coexistence (in films of finite thickness).
Hatched area indicates the possibility of conventional wetting transitions
at lower temperatures.
}
\label{fig:wetschem}
}

% ------------------- Paper heading -------------------------
%

\title{Wetting of a symmetrical binary fluid mixture on a wall}

\author{F. Schmid}

\address{Max-Planck Institut f\"{u}r Polymerforschung, D-55021 Mainz, Germany\\
         Fakult\"at f\"ur Physik, Universit\"at Bielefeld, D-33615 Bielefeld, Germany}

\author{N. B. Wilding}
\address{Department of Physics and Astronomy, The University of Edinburgh,\\
Edinburgh EH9 3JZ, U.K.
\\ Department of Mathematical Sciences, The University of Liverpool, \\Liverpool L69 7ZL, U.K.}

%\date{May 1995}
\setcounter{page}{0}
\maketitle 
\tighten

\begin{abstract}

We study the wetting behaviour of a symmetrical binary fluid below the
demixing temperature at a non-selective attractive wall. Although it
demixes in the bulk, a sufficiently thin liquid film remains mixed. On
approaching liquid/vapour coexistence, however, the thickness of the
liquid film increases and it may demix and then wet the substrate.  We
show that the wetting properties are determined by an interplay of the
two length scales related to the density and the composition
fluctuations. The problem is analysed within the framework of a generic
two component Ginzburg-Landau functional (appropriate for systems with
short-ranged interactions). This functional is minimized both
numerically and analytically within a piecewise parabolic potential
approximation. A number of novel surface transitions are found,
including first order demixing and prewetting, continuous demixing,  a
tricritical point connecting the two regimes, or a critical end point
beyond which the prewetting line separates a strongly and a weakly
demixed film. Our results are supported by detailed Monte Carlo
simulations of a symmetrical binary Lennard-Jones fluid at an
attractive wall.

\end{abstract}
\thispagestyle{empty}
\begin{center}
PACS numbers 68.45.Gd, 68.10.-m, 05.70.Jk, 68.15.+e
\end{center}

%
% ----------------------- Paper  ---------------------------
%

\section{Introduction} 
\label{sec:intro}

\begin{multicols}{2}

Phase transitions are well known to be influenced by geometrical
confinement \cite{gelb}. In practice, confinement is often imposed by
rigid external constraints, for example the surfaces of porous or
artificially nanostructured media. However, it can also be an inherent
feature of a system, as occurs for a liquid wetting film bound to a
solid substrate and in equilibrium with its vapour \cite{wetting}. In
such a situation the liquid is confined between the rigid substrate and
the flexible liquid-vapour interface.

The effects of confinement are particularly pronounced in the region of
critical points.  Under such conditions the system exhibits strong
order parameter fluctuations, the correlation length of which may become
comparable with the linear dimension of the confined system. When this
occurs the effects of confinement are felt not just near the confining
surfaces, but propagate throughout the system \cite{finitesize}. 

Critical fluctuation are relevant to the properties of liquid wetting
films if the liquid in question possesses an additional internal degree
of freedom.  Then the state of the liquid is described not only by its
number density, but by an additional parameter measuring the degree of
internal order. Examples are binary liquids, where the additional order
parameter is the relative concentration of species, and ferrofluids
where it is the magnetisation. In such systems the geometrical
constraint (i.e. the film thickness) can itself depend on the state of
order in the liquid film.  For example, simulation and experiment have
recently shown that critical concentration fluctuations can change the
equilibrium thickness of a wetting layer of a binary liquid---the
so-called critical Casimir effect \cite{joseph,nigel1,law}.

It transpires, however, that the interesting consequences of interplay
between the degree of order in the wetting layer and film thickness are
not limited to the critical point itself. To illustrate this, it is
instructive to  consider the following {\em Gedanken} experiment. Let
us take a symmetrical binary fluid, i.e. a fluid in which particles of
the same species have one strength of interaction, while interactions
between dissimilar species have another strength. As elucidated in
ref.~\cite{nigel2}, it is possible to arrange for such a system to
exhibit a line of continuous demixing transitions, terminating in a
critical end point on the liquid side of liquid-vapour coexistence.
Suppose now that the fluid is placed in contact with a non-selective
attractive substrate (wall) acting equally on both species. If the wall
is sufficiently attractive, complete wetting occurs at and above the
critical end point temperature $T_{cep}$ as liquid-vapour coexistence
is approached from the vapour side. But what happens for $T<T_{cep}$?
Far from coexistence, the wetting film is sufficiently thin that
demixing will certainly be suppressed. On approaching the coexistence
curve, however, the film 
\end{multicols}\twocolumn\noindent
 thickness grows and it is tempting to argue
that it eventually exceeds the correlation length of composition
fluctuations, whereupon the film spontaneously demixes.

Notwithstanding the appealing simplicity of this argument,   it turns
out to contain two flaws which render the actual situation rather more
complex. First, the thickness of the {\em mixed} wetting film will not
increase beyond all limits below the critical end point $T_{cep}$. This
is because a hypothetical mixed bulk liquid would not coexist with the
vapour phase at the same chemical potential $\mu_0$ as a demixed
liquid, but rather at a chemical potential which is shifted by 

\begin{equation} \mu_* - \mu_0
\propto (T_{cep} - T)^{2-\alpha} 
\end{equation} 
towards the liquid side in the $\mu-T$ plane\cite{fisher1,nigel3}. 
Since the thickness of the mixed film grows as  $\ln (\mu_* -
\mu)$\cite{wetting,dietrich1}, it is bounded from above by

\begin{equation} 
\label{l-mix} l_* \propto \ln(\mu_* - \mu_0) \propto \ln (T_{cep} - T). 
\end{equation} 
The maximum thickness $l_*$ of a hypothetical mixed film thus diverges 
logarithmically on approaching $T_{cep}$. In contrast, the correlation 
length $\xi$ diverges much faster, like $\xi \propto
(T_{cep}-T)^{-\nu}$ and will thus always exceed $l_*$ sufficiently
close to $T_{cep}$. 

The second flaw in our argument is its implicit assumption that the
composition or order parameter profile is confined in an effectively
steplike density profile, i.e. that the interfacial width between the
liquid and the vapour is much smaller than the correlation length of
composition fluctuations.  Although this is true in the region of the
critical end point, for temperature sufficiently below $T_{cep}$ the
correlation lengths of density and composition fluctuations can be
comparable and the interplay between the two subtle.

In the present study, we deploy mean field calculation and Monte Carlo
simulation to elucidate the range of possible wetting behaviour of a
symmetrical binary fluid mixture at a non-selective attractive wall for
temperatures below $T_{cep}$.  For the sake of simplicity, we have
chosen to ignore long range dispersion forces in the analytical
calculations, instead taking the interactions to be short ranged. This
allows us to base our study on a generic Ginzburg-Landau model which we
solve numerically and analytically within a square gradient
approximation. The latter leads to the construction of a film free
energy (effective interface potential) highlighting the role of the
different length scales involved in the problem. We show that the
competition of length scales results in wetting phase behavior considerably
more complex than has hitherto been appreciated. The analytical results
are compared with (and supported by) detailed Monte Carlo simulations of
a binary Lennard-Jones fluid in a semi-infinite geometry, interacting
with a non-selective attractive substrate via dispersion forces.

With regard to previous related work, the sole discussion of wetting of
symmetrical binary fluids at a non-selective wall (of which we are
aware) is that of Dietrich and Schick \cite{dietrich3} who considered
them in a sharp kink approximation treatment of binary fluids having
long-ranged interactions. Most other work on the wetting properties of
binary fluids has focused on the case of a selective substrate
(favouring one component)
\cite{dietrich1,dietrich3,hadji,kierlik,fan1}. Although such models
correspond more closely than ours to experimental conditions
\cite{dietrich1,toni}, they lack the aspect of simultaneous
demixing/ordering and wetting which is of interest to us here. It
should be stressed, however, that realisations of fluids having
symmetrical internal degrees of freedom do in fact exist, notably in
the form of ferrofluids \cite{ferrofluids}, so our model is of more
than purely theoretical interest. 

More general studies of wetting in systems with more than one order
parameter and associated length scales have been discussed by Hauge
\cite{hauge}, who pointed out that wetting exponents may become
nonuniversal even on the mean field level due to the competition of
length scales. Later studies have often focused on this
nonuniversality, e.g. in the context of wetting phenomena in
superconductors \cite{leeuwen}, alloys \cite{gompper,frank} and related
systems \cite{walden}.

The present paper is organised as follows. In section~\ref{sec:landau} we
introduce our Ginzburg-Landau free energy functional and  obtain its
wetting behaviour in the limits of infinite and vanishing order
parameter stiffness. At intermediate values of the stiffness parameter
the wetting behaviour is found firstly via an analytical minimisation of
the functional within a piecewise parabolic potential approximation
(sec.~\ref{sec:anal}), and then (in sec.~\ref{sec:num}) via a numerical
minimisation of the free energy functional to obtain the density/order
parameter profiles. In section~\ref{sec:mc} we report the results of
grand canonical Monte Carlo studies of a symmetrical binary
Lennard-Jones fluid at an attractive structureless wall. The density and
order parameter profiles with respect to the wall are obtained along a
sub-critical isotherm for a number of different wall-fluid potential
strengths. Finally we compare and discuss the mean-field and simulation
results in section~\ref{sec:concl}.

\section{Ginzburg-Landau theory}
\label{sec:landau}

Our theoretical studies are based on a generic Ginzburg-Landau functional 
for a system with two order parameters $\phi(\vec{r},z)$ and $m(\vec{r},z)$:
\begin{eqnarray}
{\cal F} &=& \int \! d \vec{r} \int_0^{\infty} \!\! dz \Big\{
\frac{g}{2} (\nabla \phi)^2 +
\frac{\gamma}{2} (\nabla m)^2 + f(m,\phi) \Big\} \nonumber \\
\label{eqn:landau}
&& \qquad
+ \int \! d\vec{r} f_s(m,\phi)|_{z=0} 
\end{eqnarray}
with the bulk free energy density
\begin{eqnarray}
f(m,\phi) & = &
-\frac{a_{\phi}}{2} \phi^2 + \frac{b_{\phi}}{4} \phi^4
-\frac{a_m}{2} m^2 + \frac{b_m}{4} m^4 \nonumber\\
\label{eqn:fb0}
&& + \mu \phi - \kappa m^2 \phi
\end{eqnarray}
and the bare surface free energy at the wall
\begin{equation}
\label{eqn:fs}
f_s(m,\phi) = \frac{C_{\phi}}{2} \phi^2 + H_{\phi} \phi
+ \frac{C_m}{2} m^2.
\end{equation}
The $z$-axis is taken to be perpendicular to the wall and
$\int d\vec{r}$ integrates over the remaining spatial dimensions.
In our case, the quantity $m$ is related to the difference between 
the partial densities of the two components, $m \propto (\rho_A - \rho_B)$,
and $\phi$ to the total density, $\phi \propto (\rho - \rho_0)$, where the 
reference density $\rho_0$ is chosen in the liquid/vapour coexistence region 
such the cubic term proportional to $(\rho-\rho_0)^3$ in (\ref{eqn:fb0}) 
vanishes. Below the liquid/vapour critical point, it is convenient to
set the units of $\phi$, $m$, $F$ and of the length such that 
$b_{m}=b_{\phi}=a_{\phi}=g=1$, and to define 
$\theta = a_m-1$. The bulk free energy density then takes the form
\begin{eqnarray}
f(m,\phi) & =& -\frac{1}{2} \phi^2 + \frac{1}{4} \phi^4
-\frac{\theta}{2} m^2 + \frac{1}{4} m^4 \nonumber\\
\label{eqn:fb}
&& - \mu \phi + \kappa (1-\phi) m^2 .
\end{eqnarray}
The bulk properties of this model have been discussed earlier\cite{nigel2}. 
A $\lambda$-line $\theta_{\lambda}(\mu)$ of continuous transitions separates 
the mixed fluid from the demixed fluid at large negative $\mu$,
corresponding to large densities $\phi$. As long as $\kappa < 1$, it is
terminated by the onset of liquid/vapour coexistence at a critical end point 
($\theta_{cep} = 0, \mu_{cep}=0$). The parameter $\mu$ is field like 
and $\theta$ is temperature like, $\theta \propto (T-T_{cep})$, 
where $T_{cep}$ is the critical end point temperature. 
Above $\theta_{cep}$, liquid/vapour coexistence is encountered at 
$\mu = 0$, and below $\theta_{cep}$, at
\begin{equation}
\label{eqn:mucoex}
\mu_c = \frac{\theta^2}{8 (1-\kappa^2)}.
\end{equation}
The coexisting liquid and gas phases are characterized by the order
parameters (to linear order in $\mu$)
\begin{equation}
\label{eqn:gas}
m^*_{-} = 0 \qquad \phi^*_{-} = -1-\mu/2
\end{equation}
in the gas phase, and
\begin{equation}
\label{eqn:liquid}
m^*_{+} = \frac{\theta-\kappa \mu}{1-\kappa^2} \qquad 
\phi^*_{+} = 1-\frac{\mu}{2} + \frac{\kappa}{2} \: m^*_{+}{}^2 
\end{equation}
in the liquid phase. These expressions are also valid in the regime where
the liquid or gas phase are metastable.

Minimizing the functional (\ref{eqn:landau}) yields the Euler Lagrange
equations 
\begin{equation}
g \frac{d^2 \phi}{(dz)^2} = \frac{\partial f}{\partial \phi}
\qquad
\gamma \frac{d^2 m}{(dz)^2} = \frac{\partial f}{\partial m}
\end{equation}
with the boundary conditions
\begin{equation}
g \frac{d \phi}{dz}\Big|_{z=0} = \frac{\partial f_s}{\partial \phi}
\qquad
\gamma \frac{d m}{d}\Big|_{z=0} = \frac{\partial f_s}{\partial m}.
\end{equation}
We wish to study a situation where the mixed liquid ($m \equiv 0$) wets
the wall at $\mu=0$ (coexistence between vapour and mixed liquid). To ensure 
this under all circumstances, we choose $H_{\phi}=-\phi_0 C_{\phi}$ with 
$\phi_0 > \phi_{+}^*$ and take the limit $C_{\phi} \to \infty$, which
is equivalent to constraining the surface density at the fixed value
$\phi(0) = \phi_0$. The surface coupling $C_m$ is taken to be positive.
It accounts for weakening of the demixing tendency at the surface due
to the reduced number of interacting neighbors.

One possible solution of the Euler Lagrange equations describes a mixed film 
at a wall. In this case, $m(z) = 0$ everywhere and one is left with
one order parameter $\phi$ only. The bulk value of $\phi$ in
the metastable mixed phase is $\phi^{(0)}_{+} = 1-\mu/2$. The standard 
way of solving the problem \cite{wetting} shall be sketched briefly
for future reference. One begins by identifying the integration constant
\begin{equation}
\label{eqn:int}
\frac{1}{2} \: (\frac{d \phi}{dz})^2 - f(\phi) + f(\phi_{-}^{(0)}) \equiv 0.,
\end{equation}
which gives an expression for $d \phi/dz$ as a function of $\phi$. 
The surface free energy can then be expressed as an integral over
$\phi$
\begin{equation}
\label{eqn:free}
F_{exc}^{(0)} = \int_{\phi_{-}^{(0)}}^{\phi_0} 
d \phi \sqrt{2 (f(\phi)-f(\phi_{-}^{(0)}))} 
%+ f_s(m_0,\phi_0),
\end{equation}
and the excess density
$\phi_{exc}^{(0)} = \frac{1}{2} \int_0^{\infty} dz 
\big[ \phi(z)-\phi_{-}^{(0)} \big]$
at the surface can be calculated {\em via}
\begin{equation}
\phi_{exc}^{(0)} = \frac{1}{2} \int_{\phi_{-}^{(0)}}^{\phi_0} 
\frac{d\phi \: (\phi - \phi_{-}^{(0)})}{\sqrt{2 (f(\phi)-f(\phi_{-}^{(0)}))}}.
\nonumber
\end{equation}
As long as $|f(\phi^{(0)}_{+} - f(\phi_{-}) | \ll
(\phi^{(0)}_{+}-\phi_{-})^2 f''(\phi^{(0)}_{+}$, which is true for
$\mu \ll 2$, the main contribution to this integral stems
from $\phi$ values around $\phi_{+}^{(0)}$. The numerator in the 
integrand can then be expanded around $\phi_{+}^{(0)}$. Carrying this
to second order and assuming 
%Up to second order in $\mu$, the term in the square root is equal to
%\begin{displaymath}
%(f(\phi)-f(\phi_{-}^{(0)})) \approx 
%\frac{1}{4}
%(\phi - \phi_-^{(0)})^2 ((\phi + \phi_-^{(0)})^2 + 2 \mu).
%\end{displaymath}
%Assuming 
$\mu \ll (\phi_0-\phi^{(0)}_{+})$, one obtains
%to yield
%\begin{displaymath}
%\phi_{exc}^{(0)} \approx 
%\int_{\phi_{-} - \phi^{(0)}_{+}}^{\phi_0-\phi^{(0)}_{+}}
%\frac{d \tau}{2} \frac{\phi^{(0)}_{+}-\phi_{-}}
%{\sqrt{2 (f(\phi^{(0)}_{+})-f(\phi_{-}))+f''(\phi^{(0)}_{-}) \tau^2}}.
%\end{displaymath}
%The solution is
\begin{eqnarray}
%\phi_{exc}^{(0)} &\approx& \sqrt{\frac{1}{8}} \ln (\frac{8}{\mu}) 
%\mu \gg (\phi_0-\phi^{(0)}_{+}) \\
\phi_{exc}^{(0)} &\approx &
\sqrt{\frac{1}{2}} \ln (\frac{2}{\phi^{(0)}_{+}-\phi_0}) 
\quad \qquad (\phi_0 < \phi^{(0)}_{+}) \\
\phi_{exc}^{(0)} &\approx&
\sqrt{\frac{1}{2}} \ln (\frac{4(\phi_0 - \phi^{(0)}_{+})}{\mu}) 
\qquad (\phi_0 > \phi^{(0)}_{+}).
\end{eqnarray}
Above the bulk demixing transition, the mixed film thus wets the
wall at coexistence ($\mu \to 0$) for $\phi_0 > \phi^{(0)}_{+}$,
and maintains a finite thickness for $\phi_0 < \phi^{(0)}_{+}$. 
We will choose $\phi_0 > \phi^{(0)}_{+}$ hereafter. From eqn. 
\ref{eqn:free}, one calculates the surface free energy to
leading order in $\mu$ 
and $(\phi_0-\phi_+^{(0)})$.
\begin{equation}
\label{eqn:fdemix}
F_{exc}^{(0)} =  \frac{2 \sqrt{2}}{3} 
 + \frac{1}{\sqrt{2}} \: (\phi_0-\phi_+^{(0)})^2 
%+ \frac{\delta \phi^{(0)}_0 {}^2}{\sqrt{2}}
%+ \frac{\delta \phi^{(0)}_0 {}^3}{3 \sqrt{2}}
%+ 2 \mu (\phi_{exc}^{(0)} + \frac{1}{\sqrt{2}})
\end{equation}
%with $\delta \phi^{(0)}_0= (\phi_0-\phi_+^{(0)})$.
Below the bulk demixing transition, $\mu = \mu_c > 0$ at coexistence
and the thickness of the mixed film remains finite under all
circumstances. 

In the following, we shall first analyse the wetting behavior 
for the limiting cases where the order parameter varies varies on very short 
length scales ($\gamma/{m_+^*}^2 \to 0$) and on long length scales 
($\gamma/{m_+^*}^2 \to \infty$) compared to the density.
Then we will discuss the general case of intermediate $\gamma$, first 
analytically in an approximation where the potential (\ref{eqn:fb}) is 
replaced by a piecewise quadratic potential, and then numerically with 
the full potential (\ref{eqn:fb}).

\subsection{Limiting cases}
\label{sec:limit}

We consider first the wetting behavior at ($\gamma/{m_+^*}^2 \to 0$). 
In this case, $m$ adapts locally to $\phi$, and the order parameter 
profile $m(z)$ can be written as $m(\phi(z))$ with
$m(\phi) = \theta + 2 \kappa (\phi-1)$ for $\phi < 1 - \theta/2 \kappa$
and $m(\phi) = 0$ otherwise. Hence we are left with the effective one
order parameter problem of calculating the density profile $\phi(z)$ in
the slightly altered potential $\hat{f}(\phi) = f(m(\phi),\phi)$. 
Since $\hat{f}(\phi)$ is a smooth function with two minima, 
one can proceed as sketched above for the mixed film, with the
analogous result: The demixed film wets the wall at 
$\phi > \phi_{+}^*$.

The analysis of the opposite case, ($\gamma/{m_+^*}^2 \to \infty$), is 
somewhat more involved. Here $\phi$ adapts locally to $m$; however, the bulk 
equation $\partial f/\partial \phi = \phi^3-\phi + \mu - \kappa m^2 = 0$
has two solutions $\phi_{\pm}(m)$. One conveniently separates the
profiles into four parts (I) -- (IV) as indicated in Figure 
\ref{fig:profschem}. The regions (I) and (III) are narrow slabs where 
$\phi(z)$ varies rapidly and $m$ can be approximated by a constant,
$m=m_0$ at the surface (I) and $m_2$ at the interface (III). 
In (I), $\phi$ drops from it's surface value $\phi_0$ to the local 
equilibrium value $\phi_{+}(m_0)$, and in (III), it switches from 
$\phi_{+}(m_2)$ to $\phi_{-}(m_2)$. The other two regions, (II) and (IV), 
are much wider; The order parameter $m(z)$ varies slowly and $\phi(z)$ 
adjusts locally to $m(z)$, such that $\phi = \phi_{+}(m)$ in (II) and 
$\phi = \phi_{-}(m)$ in (IV). 

To make the argument more quantitative, we specify the actual
subdivision of the excess free energy of (\ref{eqn:landau}),
\begin{equation}
{\cal F}_{exc} = f_s(m_0,\phi_0) 
+ {\cal F}_{I} + {\cal F}_{II} + {\cal F}_{III} + {\cal F}_{IV}
\end{equation}
with

\begin{figure}[t]
\noindent
\fig{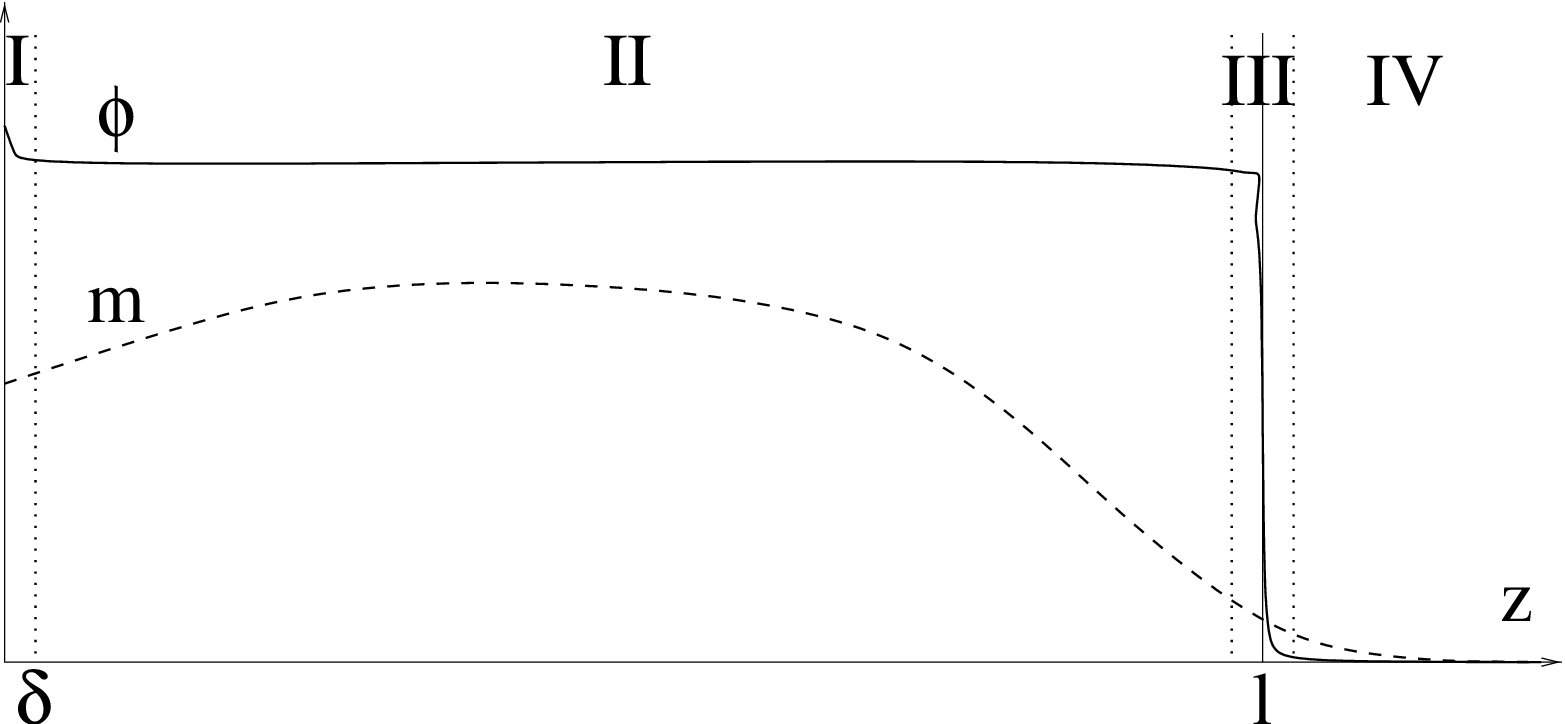}{75}{45}{
\vspace*{0.2cm}
\CCprofschem
}
\end{figure}

\begin{eqnarray*}
{\cal F}_{I} &=&
\int_0^{\delta} dz \big[ \frac{1}{2} (\frac{d \phi}{dz})^2 +
f(m_0,\phi) - f(m_0,\phi_{+}(m_0)) \big] \\
{\cal F}_{II} &=&
\int_0^l dz \big[ \frac{1}{2} \gamma (\frac{d m}{dz})^2 + 
f(m,\phi_+(m)) \big] \\
{\cal F}_{III} &=&
\int_{l-\delta}^{l} dz \big[ \frac{1}{2} (\frac{d \phi}{dz})^2 +
f(m_2,\phi) - f(m_2,\phi_{+}(m_2)) \big] \\
&+&
 \int_{l}^{l+\delta} dz \big[ \frac{1}{2} (\frac{d \phi}{dz})^2 +
f(m_2,\phi) - f(m_2,\phi_{-}(m_2)) \big] \\
{\cal F}_{IV} &=&
\int_0^l dz \big[ \frac{1}{2} \gamma (\frac{d m}{dz})^2 + 
f(m,\phi_-(m)) \big].
\end{eqnarray*}
The calculation for the regions (I), (III), and (IV) can proceed in an 
analogous way as sketched earlier for the mixed film:
The profiles of $\phi(z)$ in (I), (III), and of $m(z)$ in (IV) are monotonic 
and the integration constant (cf. \ref{eqn:int}) is known (zero).
One obtains to leading order in $\mu$ and 
$(\phi_0 - \phi_+(m_0))$
\begin{eqnarray}
F_{I} &=&
\sqrt{\frac{1}{2}} 
(\phi_0 - \phi_+(m_0))^2 
+ \cdots \\
F_{III} &=&
\frac{2 \sqrt{2}}{3} + {\cal O}((\mu - \kappa m_2{}^2)^2) \\
F_{IV} &=&
\frac{m_2{}^2}{2} \sqrt{\gamma (4 \kappa + \kappa \mu - \theta)}
\end{eqnarray}
In the region (II), the integration constant is unknown, 
\begin{equation}
\frac{1}{2} \gamma (\frac{dm}{dz})^2 - f(m, \phi_+(m)) = p,
\end{equation}
with $p>0$ if the profile of $m(z)$ is monotonic, and $p<0$ if 
$m(z)$ is nonmonotonic, like in Fig. \ref{fig:profschem}.
A connection between $p$ and the width $l$ of the film can be
established using $l=\int_{m_2}^{m_0} dm/|dm/dz|$ in the first case, and 
\begin{displaymath}
l = \int_{m_0}^{m_{max}} \frac{dm}{|dm/dz|} +
\int_{m_2}^{m_{max}} \frac{dm}{|dm/dz|} 
\end{displaymath}
in the second case, where $m_{max}$ solves $p = -f(m_{max},\phi_+(m_{max}))$. 
Next we expand the function $f(m, \phi_+(m))$ about
it's minimum $m_+^*$, leading to
\begin{equation}
f(m, \phi_+(m)) \approx (1-\kappa^2)m_+{}^4 
((\frac{m}{m_+^*} - 1)^2 -\frac{1}{4}) 
\end{equation}
One deduces the characteristic length scale,
\begin{equation}
\lambda = \sqrt{\frac{\gamma}{2 (1-\kappa^2)}} \: \frac{1}{m_+^*},
\end{equation}
which grows very large in the limit $\gamma/{m_+^*}^2 \to \infty$. 
The result for ${\cal F}_{II}$ can therefore be expanded in powers
of $e^{-l/\lambda}$.
%\begin{eqnarray}
%{\cal F}_{II} & \approx & 2 (\mu - \frac{m_+^*{}^4}{8}(1-\kappa^2)) l 
%+ m_+^*{}^2 \lambda (1-\kappa^2) \\
%&&
%\big\{ [(m_0 - m_+^*)^2 + (m_2 - m_+^*)^2] [1 + 2 A \: e^{-2 l/\lambda}
%+ (m_0 - m_+^*)(m_2 - m_+^*) e^{-l/\lambda} \big\}
%+ {\cal O}(e^{-3 l/\lambda}), \nonumber
%\end{eqnarray}
%where $A=-3$, if $m(z)$ is monotonic, and $A=1$, if $m$ is nonmonotonic.
After adding up all contributions (I)--(IV) and minimizing with respect
to $m_2$, the total excess free energy of the demixed film takes the form
$F_{exc} = F_{surf}(m_0,\phi_0) + F_{int} + V(l)$ with the
surface contribution
\begin{eqnarray}
F_{surf} &=&  f_s(m_0,\phi_0)
+ \sqrt{\frac{1}{2}} (\phi_0 - \phi_+(m_0))^2  \nonumber\\
&& + 8 \lambda \mu_c (1-m_0/m_+^*)^2 ,
\end{eqnarray}
the interface contribution
\begin{equation}
F_{int} = \: \frac{2 \sqrt{2}}{3} + 8 \lambda \mu_c,
\end{equation}
and a surface/interface interaction term
\begin{eqnarray}
V(l) &=& 2 (\mu - \mu_c) l 
- 32 \lambda \mu_c (1-m_0/m_+^*) e^{-l/\lambda} \nonumber\\
 && + 16 \lambda \mu_c (A-(1-m_0/m_+^*)^2 B)e^{-2l/\lambda},
\end{eqnarray}
where $B = [\lambda \sqrt{\gamma (\kappa - m_+^*{}^2 (1-\kappa^2)/4}]^{-1}$,
and $A=-3$ or $A=1$, depending on whether or not the profile $m(z)$ is
monotonic.

The result can now be discussed. At $m_0 < m_+^*$, the leading term
$e^{-l/\lambda}$ of the potential $V(l)$ is attractive, and wetting is 
not possible. At $m_0 > m_+^*$, an infinitely thick demixed film 
is metastable at coexistence. It's free energy difference to the mixed film 
$\Delta F =  F_{exc}-F_{exc}^{(0)}$ is up to third order in $m_+^*$
\begin{eqnarray}
\label{eqn:dfree}
\Delta F &= & \frac{1}{\sqrt{2}} \: \Big(
\sqrt{\gamma(1-\kappa^2)} \:
\big(\: 1+(1-{m_0}/{m_+^*})^2\big) \:
{m_+^*}^3 
\nonumber \\
&&
+ \: \big( \: C_m/\sqrt{2} - 
\kappa \: (\phi_0 - \phi_+(m_0))\: \big) \: m_0^2 \: \Big)
\end{eqnarray}
The limit $\gamma/{m_+^*}^2 \to \infty$ can be taken in two ways:
either $\gamma \to \infty$ at fixed $m_+^*$, or $m_+^* \to 0$ at
fixed $\gamma$. In the first case the first term in eqn.
(\ref{eqn:dfree}) dominates and the free energy of the demixed
film exceeds that of the mixed film: The film remains mixed and 
dewets accordingly.

The second case is more subtle. Here, the second term dominates, 
and the free energy of the mixed film may be less favorable, depending 
on the ratio of $C_m$ and $(\phi_0 - \phi_+^*)$. 
Note that the density enhancement at the surface, 
$\sqrt{2} \kappa (\phi_0 - \phi_+^*)$, acts as an additional
surface coupling, which opposes the effect of $C_m$. The parameter
$C_m$ accounts for the direct reduction of interacting neighbours
right at the surface. It is counterbalanced by the fact that the 
density $\phi_0$ close to the surface is higher than in the bulk.
If the latter effect dominates, the film demixes at the surface even 
for $m_+^* \to 0$ or $T \to T_{cep}$. 

\subsection{Analytical results in a piecewise parabolic potential}
\label{sec:anal}

At fixed $m_+^*$, we have seen that the demixed film wets the substrate 
in the limit $\gamma \to 0$, where the order parameter $m$ varies much
faster than the density $\phi$, and dewets at $\gamma \to \infty$,
where the density varies much faster than the order parameter.
Now we consider intermediate values of $\gamma$, where
the two characteristic length scales become comparable. Far from 
the critical end point, this is the usual case in a binary liquid,
since the interaction ranges responsible for liquid/gas separation and 
demixing are comparable.

In order to carry further the analytical analysis, we approximate
the free energy density $f(\phi,m)$ (\ref{eqn:fb}) by a
piecewise quadratic form 
\begin{equation}
\label{eqn:fbp}
f(\phi,m) = \frac{1}{2}
\Big( \phi-\tilde{\phi},m-\tilde{m} \Big)
\mbox{${\;\: \approx \atop {\textstyle f}}$}
\Big( 
\begin{array}{c} \phi-\tilde{\phi} \\ m-\tilde{m} \end{array}
\Big) + \sigma \tilde{\mu},
\end{equation}
with three pieces corresponding to the gas phase and the two liquid phases, 
separated by the lines
\begin{equation}
\label{eqn:phisep}
\phi_{sep}(m) = -\kappa(m^2+{m_+^*}^2- |m|m_+^*) + \tilde{\mu}/2.
\end{equation}
and $m \equiv 0$ at $\phi > \phi_{sep}(0)$. Here
$\tilde{\mu} = \mu-\mu_c$, $\sigma = -1$ for $\phi > \phi_{sep}(m)$
(gas phase), $\sigma = +1$ for $\phi < \phi_{sep}(m)$ (liquid phases), 
and the parabolae are adjusted to the leading terms
in the expansion of the functional (\ref{eqn:fb}) about its minima,
\begin{equation}
\Big( \begin{array}{c} \tilde{\phi} \\ \tilde{m} \end{array} \Big)
= \Big( \begin{array}{c} -1 \\ 0 \end{array} \Big),
\quad
\mbox{${\;\: \approx \atop {\textstyle f}}$} =
\Big( \begin{array}{cc} 2 & 0 \\ 0 & 4 \kappa \end{array} \Big), 
\end{equation}
for $\phi < \phi_{sep}(m)$ (gas phase), and
\begin{equation}
\Big( \begin{array}{c} \tilde{\phi} \\ \tilde{m} \end{array} \Big)
= \Big( \begin{array}{c} 1+{\kappa} {m_+^*}^2 /2
\\ \pm m_+^* \end{array} \Big),
\quad
\mbox{${\;\: \approx \atop {\textstyle f}}$} =
\Big( \begin{array}{cc}
2 & \mp 2\kappa m_+^* \\ \mp 2 \kappa m_+^* & 2 {m_+^*}^2
\end{array} \Big),
\end{equation}
for $\phi > \phi_{sep}(m)$ (liquid phases), where the upper sign holds
for $m >0$, the lower for $m<0$. The choice (\ref{eqn:phisep}) of $\phi_{sep}$
ensures that the potential $f(m,\phi)$ is continuous.

In such a potential, profiles of demixed films correspond to paths
in the $(\phi,m)$ space which can be separated into three parts:
(i) moving in one of the liquid regions from $(\phi_0,m_0)$ to $(\phi_1,m=0)$;
(ii) following the edge $(m \equiv 0)$ between the two liquid regions from 
     $(\phi_1,0)$ to $(\phi_{sep}(0),0)$;
(iii) moving in the gas region from $(\phi_{sep}(0),0)$ to $(-1,0)$.
On principle, a direct transition from (i) to (iii) is conceivable.
For the parameters $\phi_0$ of interest, however, such profiles turn out 
to be energetically less favorable than the profiles which have an
intermediate (ii). Profiles of mixed films have two parts (ii) and 
(iii) only. We shall denote $l_{(i)}\equiv l$, $l_{(ii)}$, and $l_{(iii)}$, 
the length of the slab spent in region (i), (ii) or (iii), respectively.

At given slab length and boundary conditions, the free energy 
in each of the slabs can be calculated exactly using
\begin{eqnarray}
\lefteqn{ \int_0^l dz \: \frac{1}{2} \big\{ (\frac{du}{dz})^2 +
\frac{u^2}{\lambda^2} \big\}} \\
&=& \frac{1}{4} \big\{ (u(0)+u(l))^2 \tanh \frac{l}{2 \lambda} 
+ (u(0)-u(l))^2 \coth \frac{l}{2 \lambda} \big\}. \nonumber 
\end{eqnarray}
The calculation is straightforward in the regimes (ii) and (iii). In (i),
the free energy functional has to be diagonalized first:
\begin{displaymath}
{\cal F}_{(i)} =  
%\int_0^{l_{(i)}} \!\! dz \Big\{
%\frac{1}{2} (\frac{d \phi}{dz})^2 +
%\frac{\gamma}{2} (\frac{dm}{dz})^2 + f_p(\phi,m) \Big\} \\
 \frac{1}{2} \int_0^{l} \!\! dz \Big\{
\Big[ (\frac{dv}{dz})^2 + \frac{v^2}{\lambda_1^2} \Big]
+ \Big[ (\frac{dw}{dz})^2 + \frac{w^2}{\lambda_2^2} \Big] \Big\}
\end{displaymath}
with
\begin{eqnarray}
\lambda_{1,2}^{-2} &=& 1+\frac{{m_+^*}^2}{\gamma} 
\mp \sqrt{\Big(\frac{{m_+^*}^2}{\gamma}\Big)^2 + \frac{{m_+^*}^2}{\gamma}
(4 \kappa^2-2)+1}, \nonumber\\
\Big( \begin{array}{c} v\\w \end{array} \Big) &=& 
\frac{1}{\sqrt{e^{\delta}+e^{-\delta}}} 
\Big( \begin{array}{cc} 
e^{-\delta/2} & e^{\delta/2}\\e^{\delta/2} &- e^{-\delta/2} 
\end{array} \Big) 
\Big( \begin{array}{c} \phi-\tilde{\phi}\\\sqrt{\gamma} (m-\tilde{m}) 
\end{array} \Big), \nonumber\\
\end{eqnarray}
where we have defined
\begin{equation}
\delta = \frac{1}{2} \ln ( \frac{2-\lambda_1^{-2}}{\lambda_2^{-2}-1}).
\end{equation}
The parameter $\delta$ or alternatively $\gamma/{m_+^*}^2$ 
determines the wetting behavior. Fig. \ref{fig:lg} shows the two length 
scales $\lambda_1$ and $\lambda_2$ as a function of $\gamma/{m_+^*}^2$. 
The length $\lambda_1$ is always larger than $\lambda_2$. 
At $\gamma/{m_+^*}^2 \gg 1$ or $\delta \gg 0$, it characterizes the spatial 
variations of $m(z)$ and grows linearly with $\gamma/{m_+^*}^2$; 
at $\gamma/{m_+^*}^2 \ll 1$  or $\delta \ll 0$, it characterizes the 
variations of $\phi(z)$  and remains largely independent of 
$\gamma/{m_+^*}^2$. These are the limiting regimes discussed in the 
previous subsection. At $\gamma/{m_+^*}^2 \approx 1$ or 
$\delta \approx 0$, both $\lambda_1$ and $\lambda_2$ are related to
linear combinations of $\phi(z)$ and $m(z)$. 

The further calculation proceeds as follows: The free energy in 
(iii) is given by
\begin{equation}
F_{(i)} = \sqrt{2}/2 (1+ (\phi_{sep}(0)+1)^2).
\end{equation}
In the region (ii), the result for the free energy is expanded in 
powers of $e^{-\sqrt{2} l_{(ii)}}$ up to the second order and minimized
with respect to $l_{(ii)}$. 
The free energy calculated in (i) is expanded up to second order in powers 
of $e^{-l/\lambda_1}$ and up to first order in $e^{-l/\lambda_2}$, where 
$l \equiv l_{(i)}$. The three contributions are then added up, and the sum 
is minimized with respect to $\phi_1$ and $m_0$ at given surface coupling
$C_m$. The solution has to be compared with the free energy of a mixed film,
which is calculated analogously.

\begin{figure}[t]
\noindent
\fig{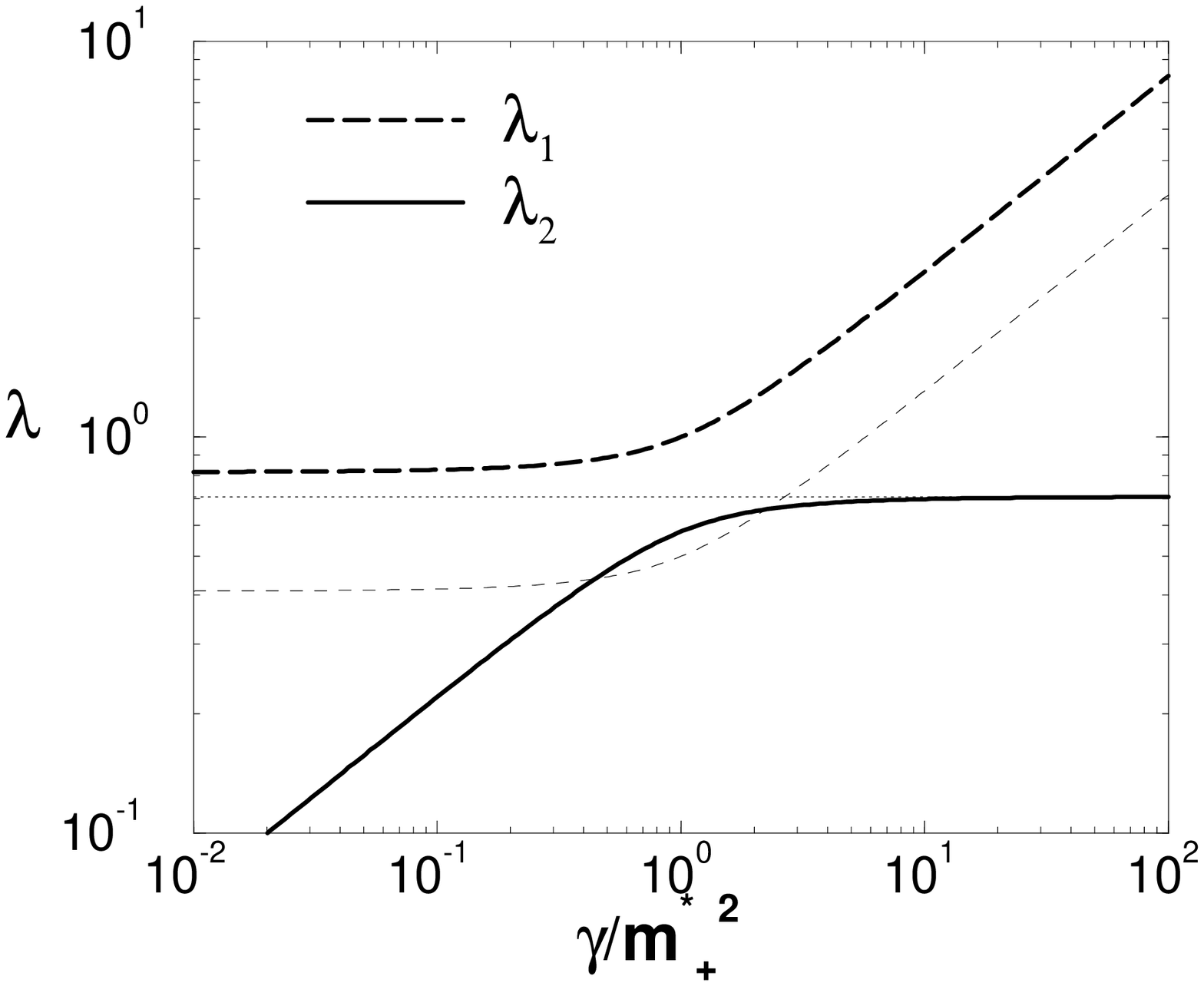}{75}{65}{
\vspace*{0.2cm}
\CClg
}
\end{figure}

We only report the result for the case $C_m=0$ here. The expressions 
obtained for arbitrary $C_m$ are more complicated, but qualitatively 
similar. Without loss of generality, we can assume $m>0$ in the 
demixed film. As long as $m_0 > 0$, the surface order parameter $m_0$
and the free energy difference $\Delta F$ between the mixed and demixed 
film can then be expanded as
\begin{eqnarray}
\sqrt{\gamma} \: m_0 &=& \sqrt{\gamma} \: m_+^* + 
 \iota_0 + \iota_1 e^{-l/\lambda_1}
+ \iota_2 e^{-l/\lambda_2} + \iota_3 e^{-2 l/\lambda_1} 
\nonumber \\
\label{eqn:expansion}
\Delta F(l) &=& \Delta F_{(ii)} + 
  \tau_0 + \tau_2 e^{-l/\lambda_1}
+ \tau_2 e^{-l/\lambda_2} + \tau_3 e^{-2 l/\lambda_1}.
\nonumber\\
\end{eqnarray}
Using the abbreviations 
\begin{displaymath}
K_0 = \frac{e^{\delta}+e^{-\delta}}{\lambda_1 \lambda_2}, 
\quad \mbox{and} \quad
K_{\pm} = \frac{\lambda_1 \lambda_2}
{e^{\pm \delta} \lambda_1 + e^{\mp \delta} \lambda_2},
\end{displaymath}
the coefficients can be written as
\begin{eqnarray}
{\iota}_0 &=& 
(\lambda_2^{-1}-\lambda_1^{-1}) K_- \cdot ({\phi}_0 - \tilde{\phi})
 \nonumber\\
{\iota}_{1,2} &=&
-2 K_0 K_+ K_- e^{\pm \delta} \cdot \sqrt{\gamma} m_+^* 
\label{eqn:iota}
\\
{\iota}_3 &=& 
-2 K_0 K_-^2 (1+8 K_+ e^{-\delta}/\lambda_1) \cdot 
({\phi}_0 -\tilde{\phi})
\nonumber
\end{eqnarray}
\begin{eqnarray}
{\tau}_0 &=&
K_0 (K_+ \cdot \gamma {m_+^*}^2 + K_- \cdot 
({\phi}_0-\tilde{\phi})^2) \nonumber\\
{\tau}_{1,2} &=& 
\pm 2 K_0 K_+ K_-/\lambda_{2,1} \cdot \sqrt{\gamma} m_+^* \cdot 
({\phi}_0 -\tilde{\phi})
\label{eqn:tau}
\\
{\tau}_3 &=&
K_0 K_- K_+^2 
e^{\delta} ({e^{-\delta}}/{\lambda_2} - {e^{\delta}}/{\lambda_1})/\lambda_2 
\cdot \gamma {m_+^*}^2 
\nonumber \\
&& + \: K_0 K_-^2 e^{-\delta} 
(1+8 K_+ e^{-\delta}/\lambda_1) /\lambda_2 \cdot 
({\phi}_0-\tilde{\phi})^2,
\nonumber
\end{eqnarray}
and with $h = {m_+^*}^4(1-\kappa^2) - 2 \tilde{\mu}$, 
$p = \phi_0 - \tilde{\phi} + \kappa {m_+^*}^2$,
\begin{equation}
\Delta F_{(ii)} = -\frac{1}{\sqrt{2}} \: p^2
- \frac{h}{2 \sqrt{2}} \Big(\: 1 + \ln \frac{4 p^2}{h} \: \Big).
\end{equation}
%\begin{eqnarray}
%\lefteqn{\Delta F_{(ii)} = -\frac{1}{\sqrt{2}} 
%(\phi_0 - \tilde{\phi} + \kappa {m_+^*}^2)^2} \\ 
%&&- \: \frac{{m_+^*}^4(1-\kappa^2) - 2 \tilde{\mu}}{2 \sqrt{2}} 
%\Big(\: 1 + \ln \frac{4 (\phi_0 - \tilde{\phi} + \kappa {m_+^*}^2)^2}
%{{m_+^*}^4(1-\kappa^2) - 2 \tilde{\mu}} \: \Big). \nonumber
%\end{eqnarray}
When taking the limits $\delta \to \pm \infty$, one recovers 
qualitatively\cite{footnote} the behavior discussed in the previous section.

\begin{figure}[t]
\noindent
\fig{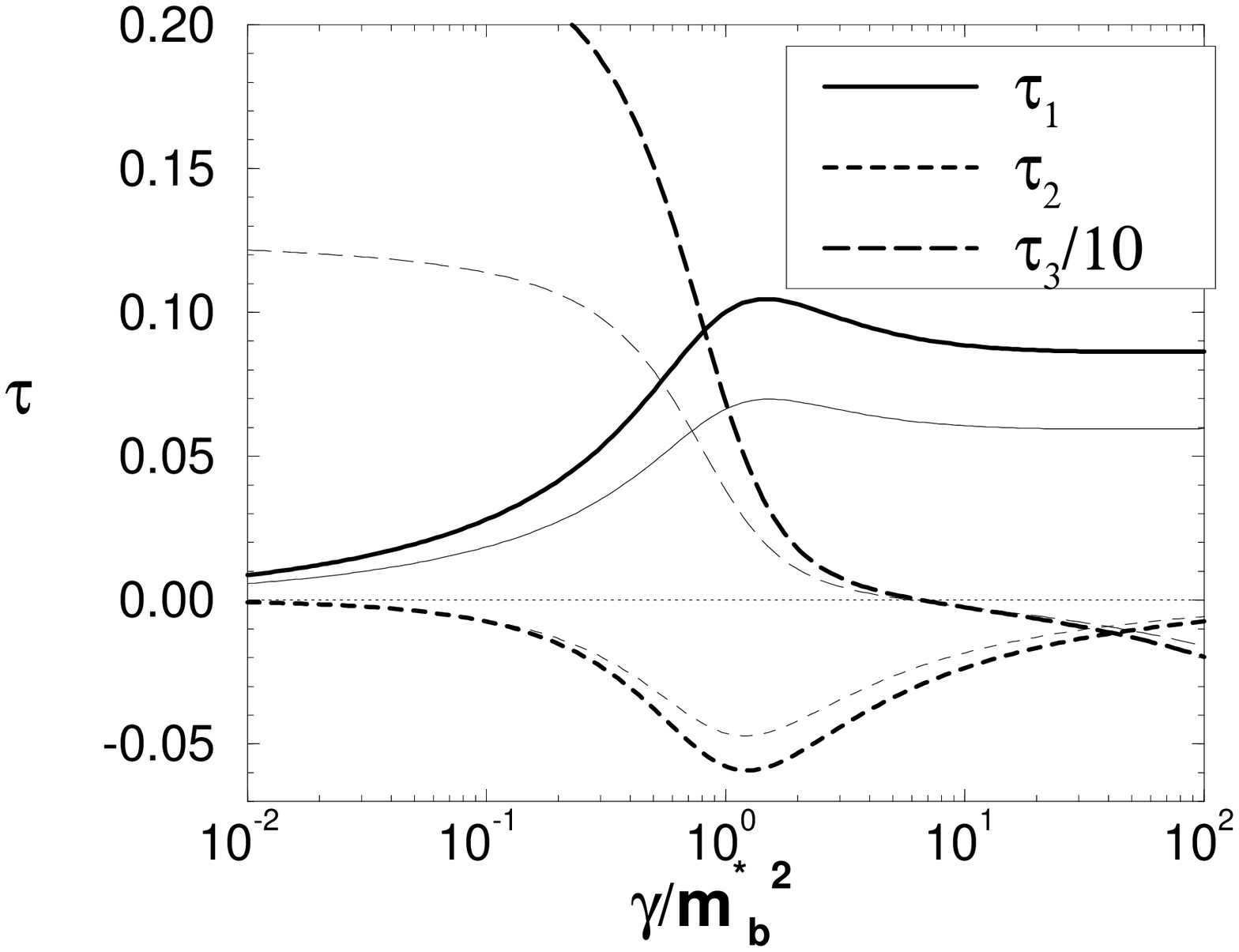}{75}{65}{
\vspace*{0.2cm}
\CCtg
}
\end{figure}

The function $\Delta F(l)$ can be conceived as an effective interface
potential for the demixed film. The parameters $\tau_i$ for a choice 
of $\phi_0$ ($\phi_0=1.5$) and two values of $C_m$,
$C_m=0$ according to (\ref{eqn:tau}) and $C_m=0.02$, are shown
in Fig. \ref{fig:tg}. One finds that $\tau_1$ is always positive, $\tau_2$ 
is always negative, and $\tau_3$ changes sign from positive to negative as
$\gamma/{m_+^*}^2$ increases. The leading term of the potential $F(l)$ is
thus positive, and one expects a first order wetting transition
and a prewetting line. On the other hand, the expansions
(\ref{eqn:expansion}) are only valid as long as the surface order 
parameter $m_0$ is positive. According to eqn. (\ref{eqn:iota}), the 
coefficients $\iota_i$ of the expansion for $m_0(l)$ are negative except 
for the zeroth order term $\iota_0$. Hence $m_0$ decreases with film 
thickness and may vanish at some thickness $l_c$. In this case, the film 
mixes continuously at $l_c$, and the prewetting line turns into a second 
order demixing line sufficiently far from coexistence. 

\subsection{Numerical solution}
\label{sec:num}

The analytical results of the previous subsection provided
insight into the competition of length scales in the binary
fluid and the wetting scenarios which can be expected on a wall
as a result. However, a reliable calculation of actual phase 
diagrams, including the details of the prewetting line, is not possible
on the basis of the expansion (\ref{eqn:expansion}). We have thus 
supplemented the analytical work by a numerical minimization of the
functional (\ref{eqn:landau}) in the $\mu-\phi_0$ plane
for selected sets of parameters $\gamma$ and $C_m$.

The problem is simplified considerably due to the
fact that $\phi(z)$ is a monotonic function of $z$, i.e.,
$m(z)$ can be expressed as a function $m(\phi)$. The bulk free 
energy functional in (\ref{eqn:landau}) can thus be rewritten as
\begin{eqnarray*}
{\cal F} &=& \int_0^{\infty} dz \big\{ \frac{1}{2} 
(1 + \gamma (\frac{dm}{d\phi})^2) \: (\frac{d \phi}{dz})^2 
+ f(m(\phi),\phi) \big\} \\
&=& \int_{\phi_-^*}^{\phi_0} d \phi \:
\sqrt{1+\gamma (\frac{dm}{d\phi})^2} \sqrt{f(m(\phi),\phi)-f(0,\phi_-^*)},
\end{eqnarray*}
where the integration constant (\ref{eqn:int}) has been identified and
exploited as usual. Minimization with respect to the function $m(\phi)$ 
leads to the Euler-Lagrange equation
\begin{equation}
2 \gamma f(m,\phi) \frac{d^2 m}{d \phi^2} =
(1 + \gamma (\frac{dm}{d\phi})^2) \: 
(\frac{\partial f}{\partial m} 
- \gamma \frac{dm}{d \phi} \: \frac{\partial f}{\partial \phi} ),
\end{equation}
which we have solved using the Verlet algorithm.

Some results are shown in Figs. \ref{fig:phwet1}, \ref{fig:profiles},
\ref{fig:phwet2}, and \ref{fig:phwet3}. As anticipated in the previous 
subsection, we find a first order wetting transition, a discontinuous 
prewetting line and a continuous demixing line. At surface coupling
$C_m>0$, the demixing line joins the prewetting line in a surface
critical end point (Fig. \ref{fig:phwet1}). The prewetting line
separates a demixed thick film  from a mixed thin film (see profiles 

\begin{figure}[t]
\noindent
\fig{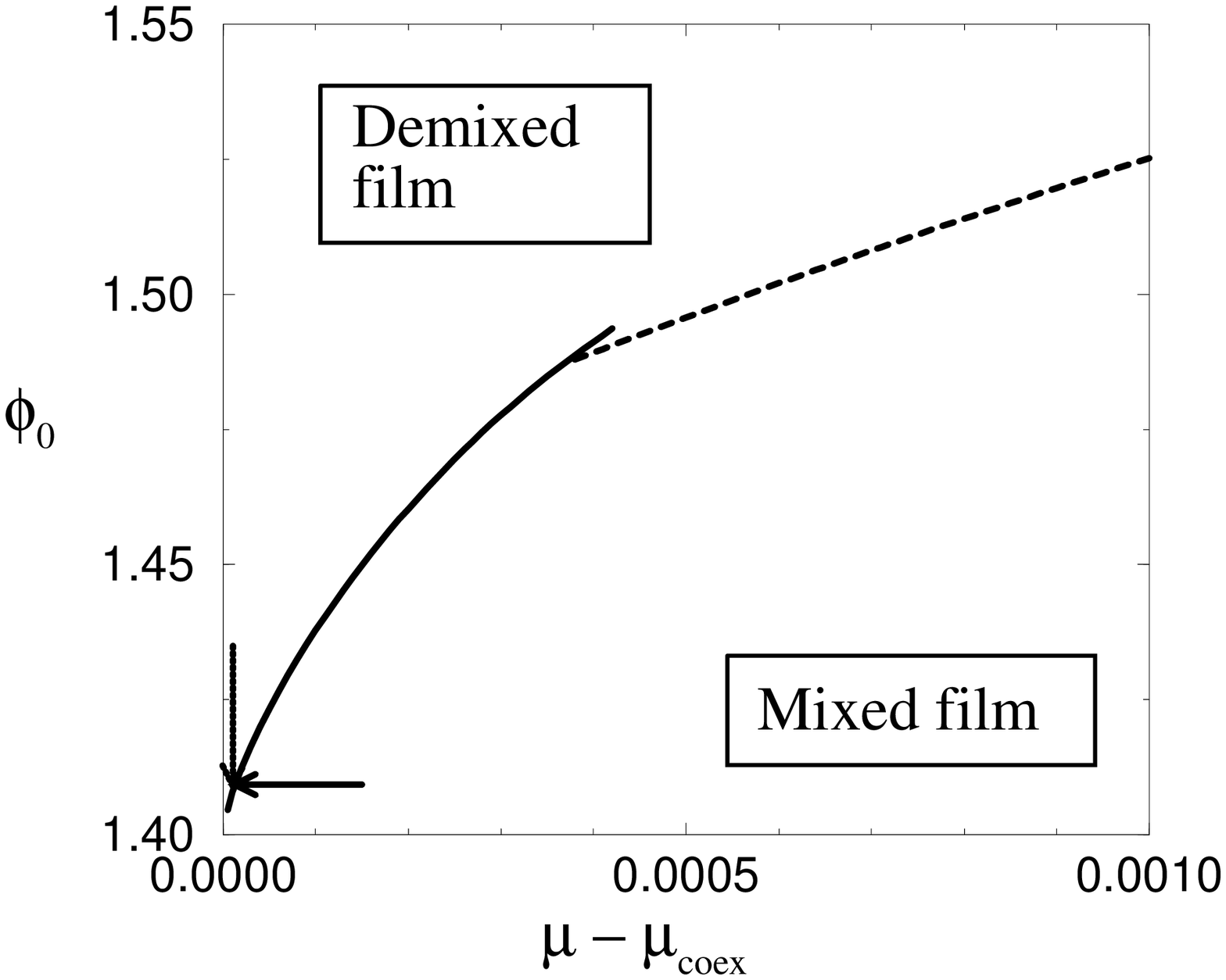}{75}{65}{
\vspace*{0.2cm}
\CCphweta
}
\end{figure}

\begin{figure}[t]
\noindent
\fig{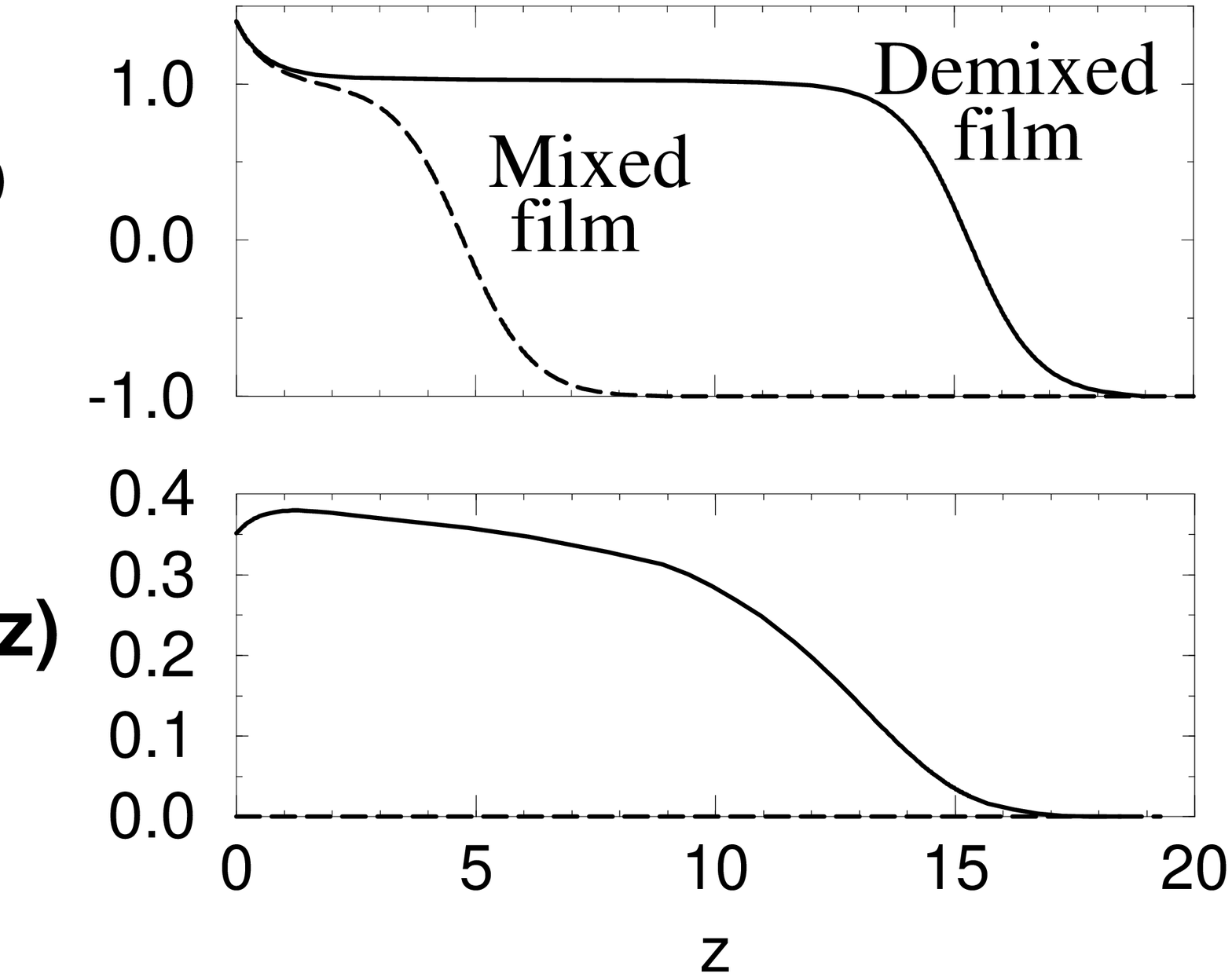}{70}{60}{
\vspace*{0.2cm}
\CCprofiles
}
\end{figure}
\noindent
in
Fig. \ref{fig:profiles}) before reaching the critical end point, then
two demixed films of  different thickness, and finally vanishes in a
critical point. On decreasing the surface coupling $C_m$, the critical
end point and  the critical point move closer to each other, until they
merge in a surface tricritical point. 

Fig. \ref{fig:phwet2} shows two cases of phase diagrams in the 
$\phi_0-\mu$ plane for $C_m=0$ and two different $\gamma$ at fixed 
$\theta$, i. e., at fixed bulk order parameter $m_+^*$.
With increasing $\gamma$, the prewetting line shifts towards larger 
$\phi_0$ and extends deeper into the off-coexistence region. 
As $\gamma \to \infty$, it moves to $\phi_0 \to \infty$, the 
film remains mixed and thin at all finite $\phi_0$. 
At $\gamma \to 0$, on the other hand, the line becomes
flat, approaches $\phi_+^*$, and the tricritical point where it turns
into a second order line moves to $\mu_t \to \mu_c$. 
The numerical results thus agree with the conclusions from 
section \ref{sec:limit}.

Fig. \ref{fig:phwet3} demonstrates what happens if instead of
making $\gamma$ larger, one increases the characteristic length scale of 
order parameter fluctuations by decreasing $\theta$, i.e., approaching
the  critical end point (reducing $m_+^*$). Far from liquid/vapour
coexistence,  the transition line still moves towards larger $\phi_0$. 
However, the effect reverses close to coexistence, the demixing 
transition is now shifted to lower surface densities $\phi_0$. 
Furthermore, the length of the prewetting line shrinks instead of
growing.

\begin{figure}[t]
\noindent
\fig{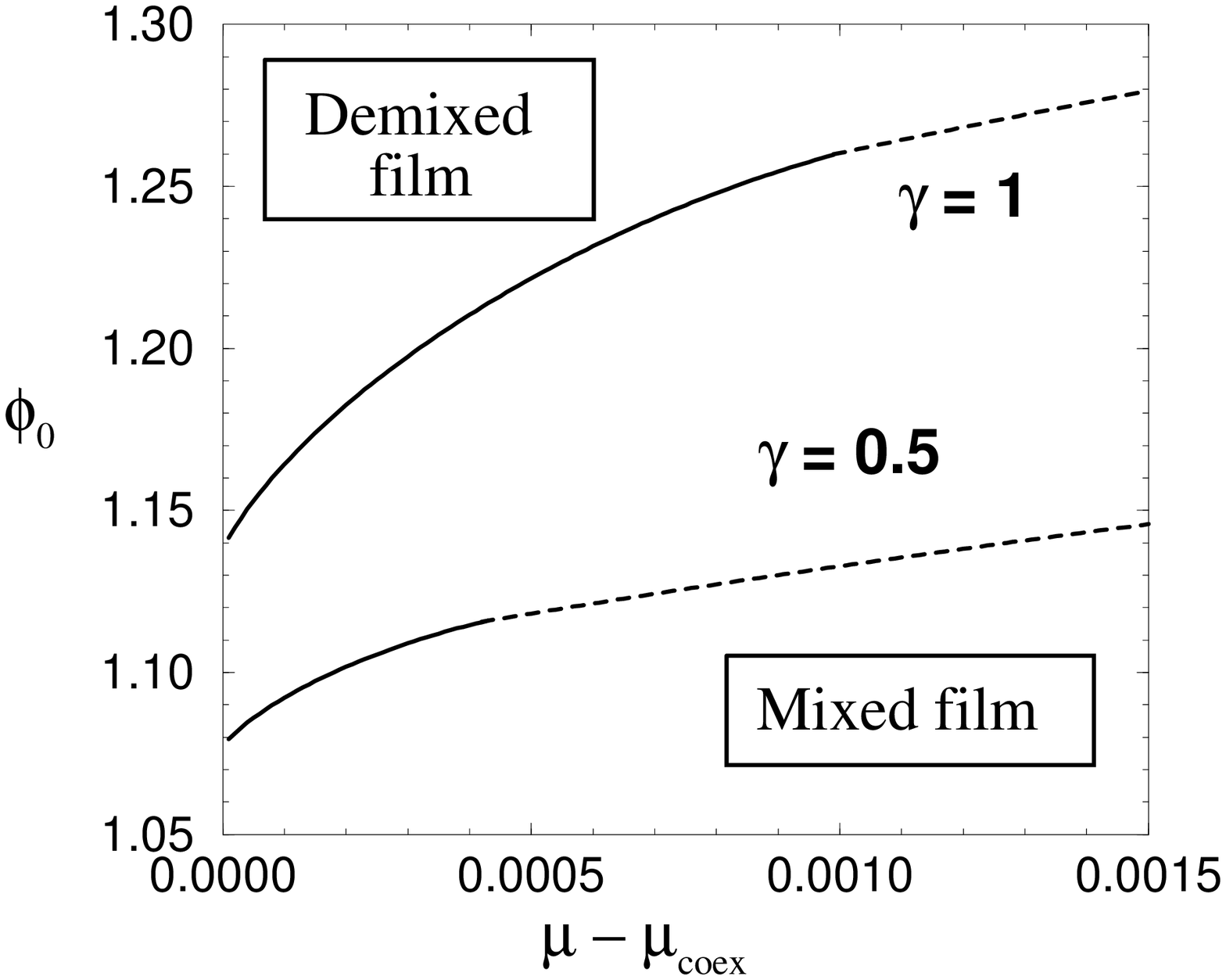}{75}{70}{
\vspace*{0.2cm}
\CCphwetb
}
\end{figure}

\begin{figure}[t]
\noindent
\fig{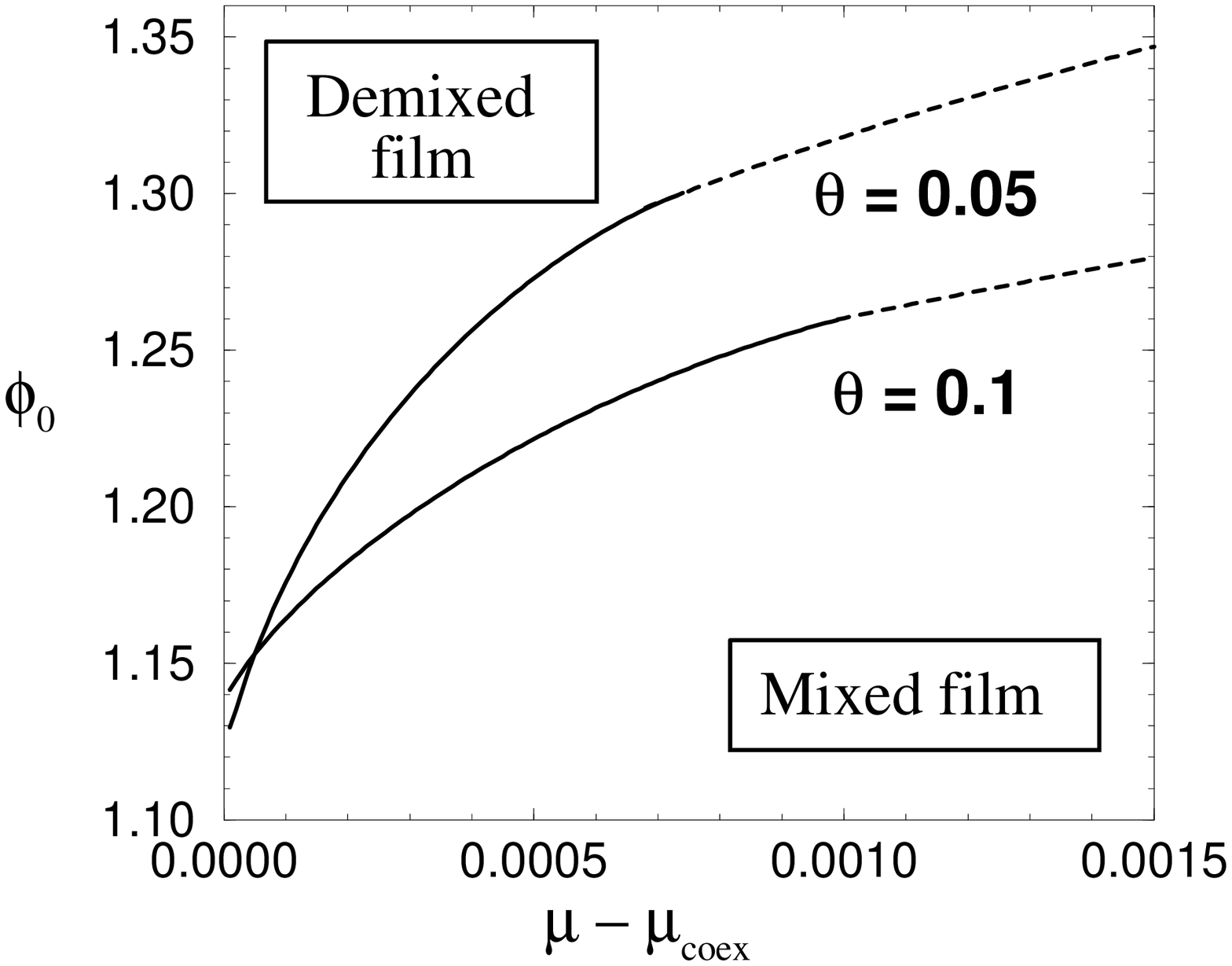}{75}{65}{
\vspace*{0.2cm}
\CCphwetc
}
\end{figure}

\section{Monte Carlo simulations}

In this section we describe Monte Carlo simulation studies of the
subcritical wetting behaviour of a symmetrical binary fluid at a
structureless wall.

\subsection{Model and simulation details}
\label{sec:mc}

The system we have studied is a symmetrical binary fluid,
having interparticle interactions of the Lennard-Jones (LJ) form:

\begin{equation}
u(r_{ij})=4\epsilon_{ij}\left[\left(\frac{\sigma_{ij}}{r_{ij}}\right)^{12}-
\left(\frac{\sigma_{ij}}{r_{ij}}\right)^6\right]
\end{equation} 

We made the following choice of model parameters: $\sigma_{11}=
\sigma_{22}= \sigma_{12}=\sigma=1$; $\epsilon_{11}= \epsilon_{22}=
\epsilon$; $\epsilon_{12}= 0.7\epsilon$. i.e. interactions between
similar species are treated identically, but those between unlike
species are weakened.  The inter-particle potential was truncated at a
distance of $R_c=2.5\sigma$ and no long-range correction or potential
shift was applied.

The fluid was confined within a cuboidal simulation cell having
dimensions $P_x\times P_y\times D$, in the $x,y$ and $z$ coordinate
directions respectively, with $P_x=P_y\equiv P$. The simulation cell
was divided into cubic sub-cells (of size the cutoff $R_c$) in order to
aid identification of particle interactions. Thus  $P=pR_c$  and
$D=dR_c$, with $p$ and $d$ both integers.  To approximate a semi-infinite
geometry, periodic boundary conditions were applied in the $x$ and $y$
directions, while hard walls were applied in the unique $z$ direction
at $z=0$ and $z=D$. The hard wall at $z=0$ was made attractive, using a
potential designed to mimic the long-ranged dispersion forces between
the wall and the fluid \cite{israel}:

\begin{equation}
V(z)=\epsilon_w\left[\frac{2}{15}\left(\frac{\sigma_w}{z}\right)^9-
\left(\frac{\sigma_w}{z}\right)^3\right]
\end{equation}
Here $z$ measures the perpendicular distance from the wall,
$\epsilon_w$ is a `well-depth' controlling the interaction strength,
and we set $\sigma_w=1$. No cutoff was employed and the wall potential
was made to act {\em equally} on both particle species.

Monte-Carlo simulations of this system were performed using a
Metropolis algorithm within the grand canonical ($\mu,V,T$) ensemble
\cite{FRENKEL}. Three types of Monte-Carlo moves were employed:

\begin{enumerate}
\item Particle displacements
\item Particle insertions and deletions
\item Particle identity swaps: $1\to 2$ or $2\to 1$
\end{enumerate}
To maintain the symmetry of the model, the chemical potentials
$\mu_1$ and $\mu_2$ of the two components were set equal at all times.
Thus only one free parameter, $\mu=\mu_1=\mu_2$, couples to the
overall number density $\rho=(N_1+N_2)/V$. The other variables used to
explore the wetting phase diagram were the reduced well depth
$\epsilon/k_BT$ and the reduced wall potential
$\epsilon_w/k_BT$. During the simulations, the observables monitored
were the total particle density profile

\begin{equation} 
\rho(z)= [N_1(z)+N_2(z)]/P^2 \;,
\end{equation} 
the number difference order parameter profile,
\begin{equation} 
n(z)= |N_1(z)-N_2(z)|/P^2
\end{equation} 
These profiles was accumulated in the form of a histogram. Other
observables monitored were the total interparticle energy and the wall
interaction energy.

The choice of system size was, as ever, a compromise between minimising
finite-size effects and maximising computational throughput.  Tests
showed the profiles to be largely insensitive to the size of the wall
area and hence $p=7$ was used, this being the largest computationally
tractable size consistent with the necessary choice of the slit width
$d$. The latter must clearly be considerably larger than the film
thicknesses of interest in order to prevent the liquid film directly
interacting with the hard wall at $z=D$. In the results presented
below, the typical slit width used was $d=16$, corresponding to some
$40$ molecular diameters. For thin films a narrower slit of width $d=8$ was 
used.

\subsection{Wetting behaviour along a subcritical isotherm}

\label{sec:res}

Accurate knowledge of bulk phase behaviour is an essential prerequisite
for detailed studies of near-coexistence wetting properties. In the
present model, the locus of the liquid vapour coexistence curve and
location
\begin{figure}[t]
\noindent
\fig{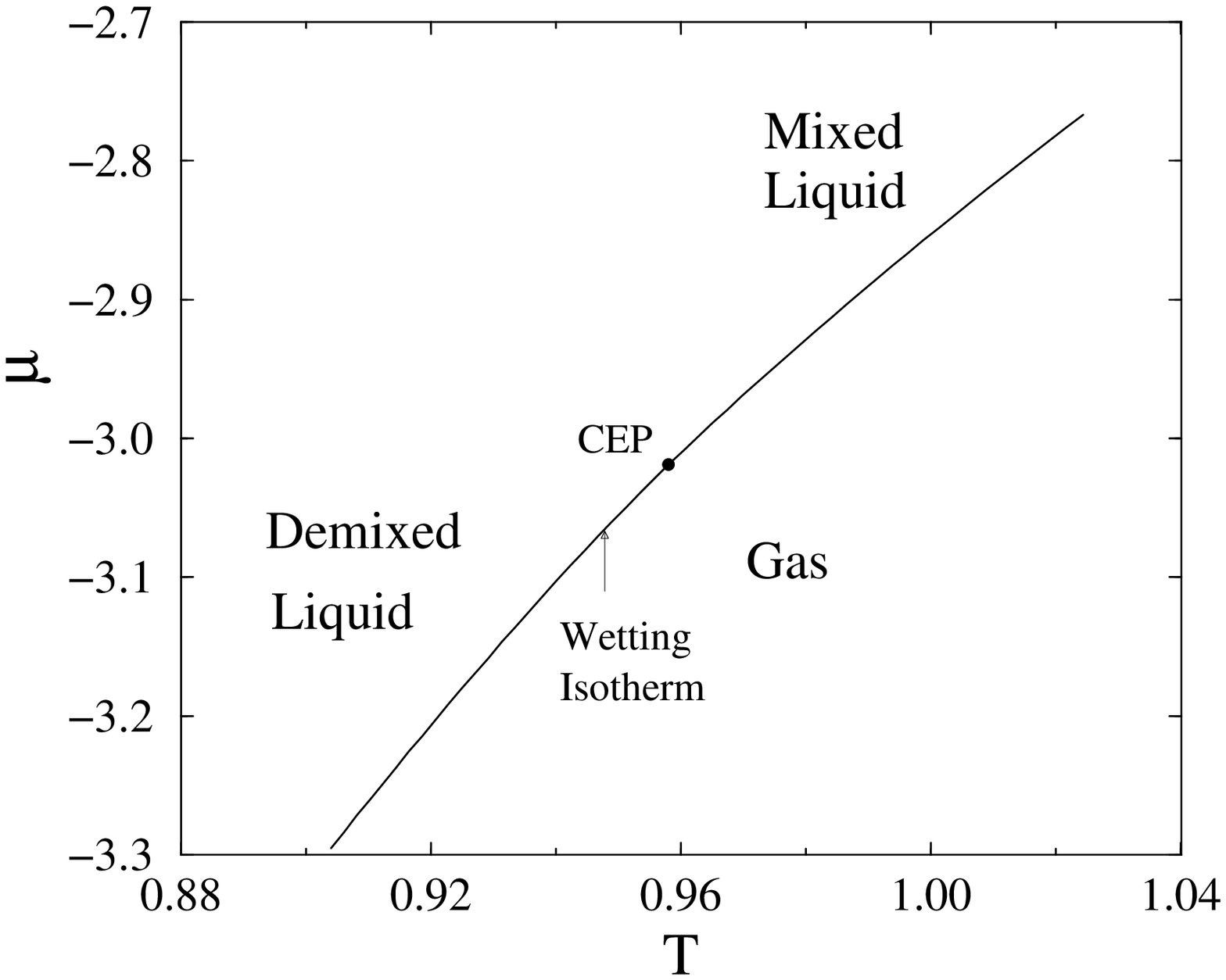}{65}{55}{
\vspace*{1.cm}
\CCmuT
}
\end{figure}
\noindent
of the critical end point are already known to high precision
from a previous MC simulation study \cite{nigel3,nigel1}.  The phase
diagram in the $\mu$-$T$ plane (in standard Lennard-Jones
reduced units \cite{FRENKEL}) is shown in fig.~\ref{fig:muT}. 
The critical end point is located at $T_{cep}=0.958(3),
\mu_{cep}=-3.017(3)$ \cite{nigel3,nigel1}. We note that although the
locus of the coexistence curve is known to five significant figures,
the position of the CEP {\em along} this tightly determined line is
known only to three significant figures. 

To determine the wetting properties at temperatures below $T_{cep}$,
the number density profile $\rho(z)$ was studied along the isotherm
$T=0.9467$ as coexistence was approached from the vapour side. To
achieve this, the chemical potential was incremented up to its
coexistence value $\mu_{cx}(T)$ in a sequence of $6$-$10$ steps of
constant size $\Delta\mu=0.0025$. This procedure was repeated for a
number of different values of the wall-fluid potential strength
$\epsilon_w$, allowing the influence of this parameter on the wetting
behaviour to be ascertained.  In all, six values of the $\epsilon_w$
were studied  ($\epsilon_w=1.0, 1.7, 1.75, 2.0, 3.0, 4.0$). We describe the
wetting behaviour for each in turn.

For $\epsilon_w=1.0$, fig.~\ref{fig:ew1.0} shows that although
the film thickness grows very slightly as coexistence is approached,
it never exceed two molecular diameters. At no point in
the profile does the density attain that of the liquid phase
($\rho\approx 0.6$). The presence of a thin wetting layer right up to
coexistence implies incomplete (partial) wetting.

Increasing the wall potential to $\epsilon_w=1.70$
[fig.~\ref{fig:ew1.7}], results in considerably more structure near the
wall compared to $\epsilon_w=1.0$, with clear density oscillations
arising from excluded-volume `packing effects' \cite{finn}. The profile
is much more responsive to changes in the chemical potential
and reaches a thickness of 4-5 molecular diameters close to
coexistence. 

For $\epsilon_w=1.75$, however, the situation changes
qualitatively, as shown in fig.~\ref{fig:ew1.75}. On increasing the
chemical potential, a clear jump is observed in both the thickness of
the film, and the value of its density. In the thick film, the density
of a significant portion of the film is that of

\begin{figure}[t]
\noindent
\fig{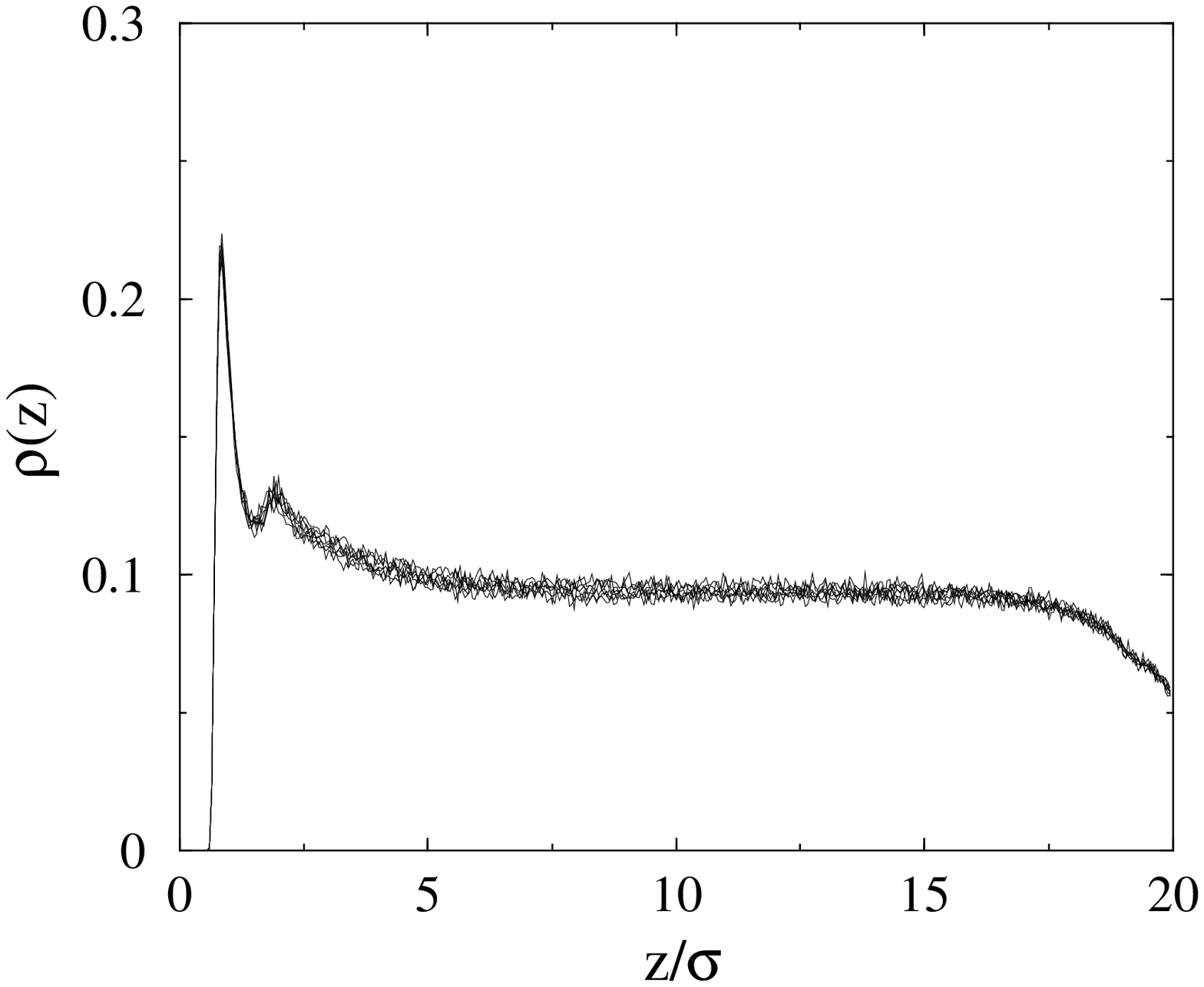}{65}{55}{
\vspace*{1.cm}
}
\CCewA
\end{figure}

\begin{figure}[t]
\noindent
\fig{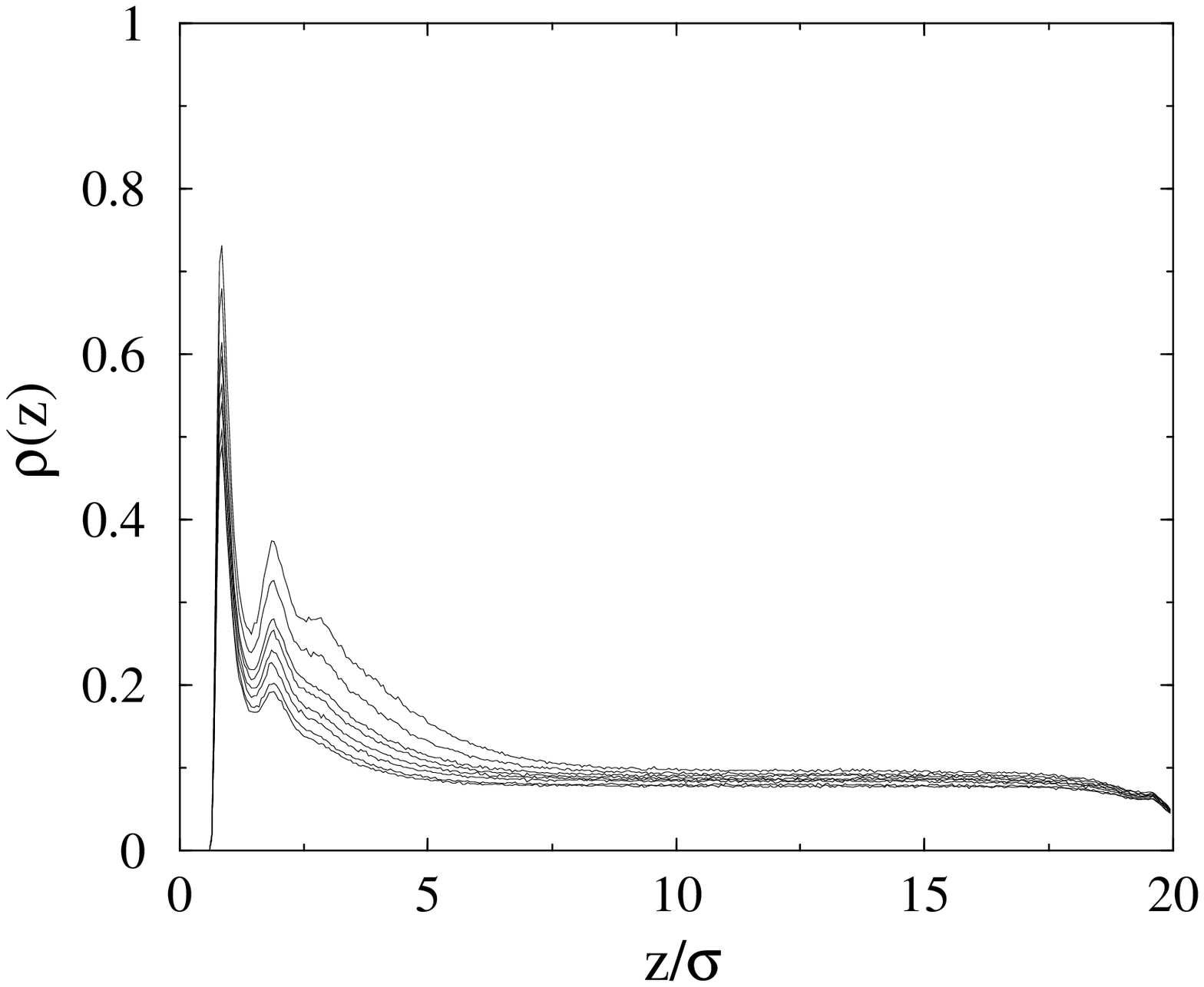}{65}{55}{
\vspace*{1.cm}
}
\CCewB
\end{figure}

\begin{figure}[t]
\noindent
\fig{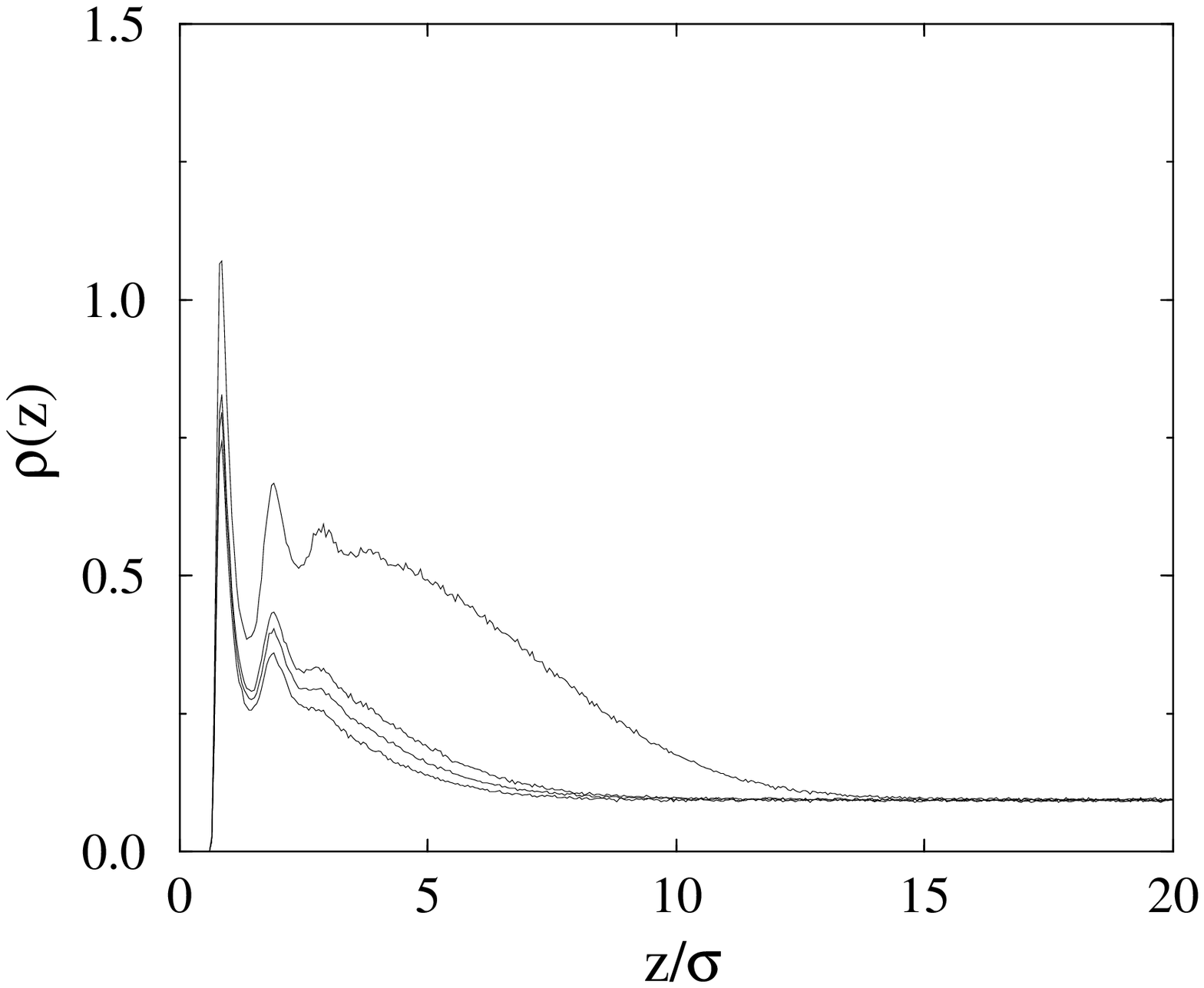}{65}{55}{
\vspace*{1.cm}
}
\CCewC
\end{figure}

\begin{figure}[t]
\noindent
\fig{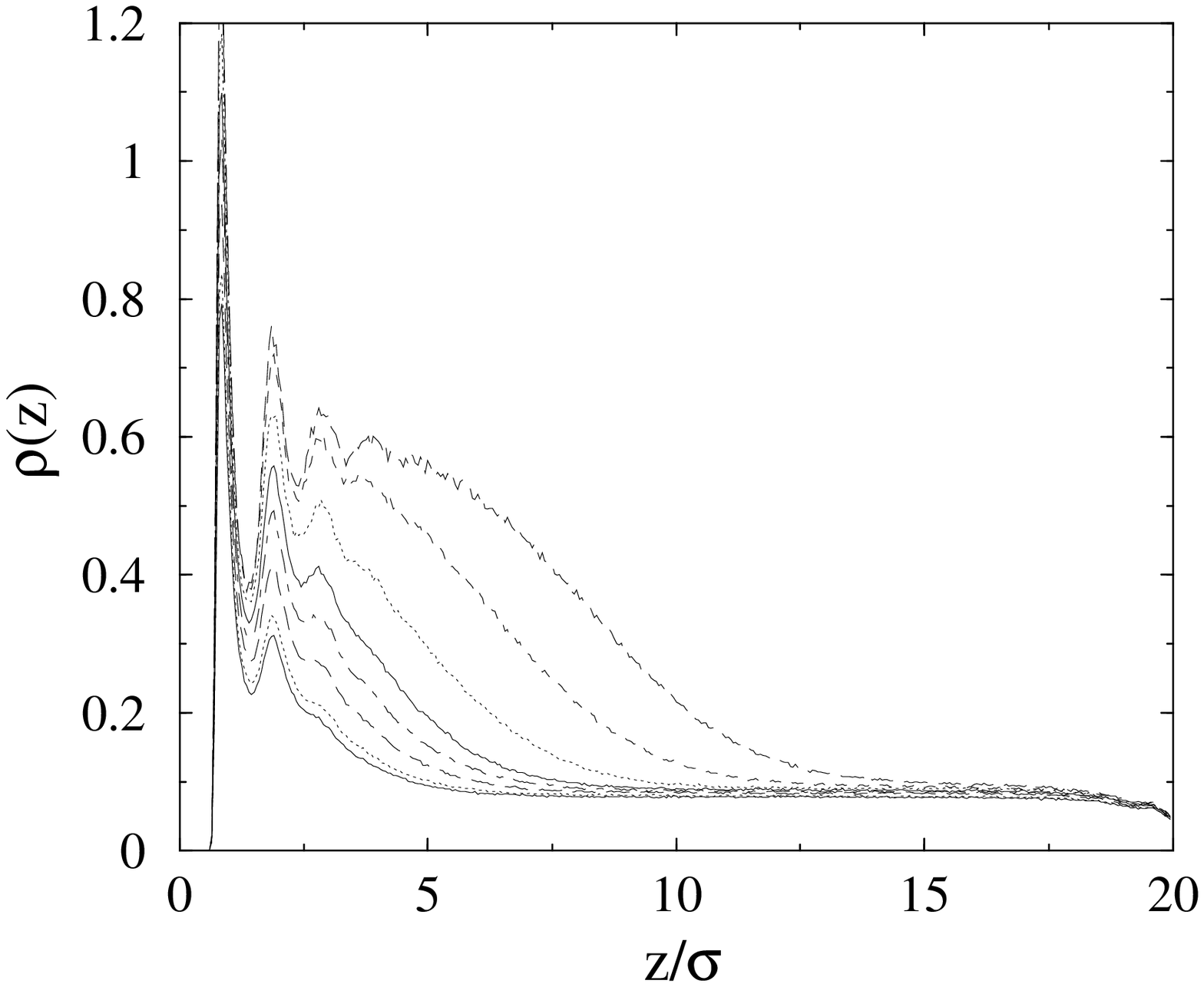}{65}{55}{
\vspace*{1.cm}
}
\CCewD
\end{figure}
\noindent
the bulk liquid. This
thin-thick jump constitutes a prewetting transition, as previously
observed in simulation studies of lattice gas models \cite{nicolaides},
Lennard-Jones fluids \cite{kierlik,fan1,finn,fan2} as well as experimentally
\cite{kellay}.

As the wall potential is increased to $\epsilon_w=2.0$
(fig.~\ref{fig:ew2}), the sharp prewetting transition is lost and
instead the film thickness increases smoothly as $\mu$ approaches its
coexistence value. This suggests that here the system is above the
prewetting critical point (\cite{nicolaides}).

On increasing $\epsilon_w$ to $3.0$, a new feature emerges
(fig.~\ref{fig:ew3}(a)). As the chemical potential increases, the
thickness of the film initially increases smoothly with increasing
$\mu$. However, once the thickness reaches some $10$ molecular
diameters, a large jump occurs to a thickness of about $15$ molecular
diameters. Concomitant with this jump is a demixing of the film as a
whole, as seen in the order parameter profile fig.~\ref{fig:ew3}(b).
The size of the jump in the layer thickness appears to reduce as the
wall strength is increased to  $\epsilon_w=4.0 $ (fig.~\ref{fig:ew4}),
suggesting a weakening of the transition.

\begin{figure}[t]
\noindent
\fig{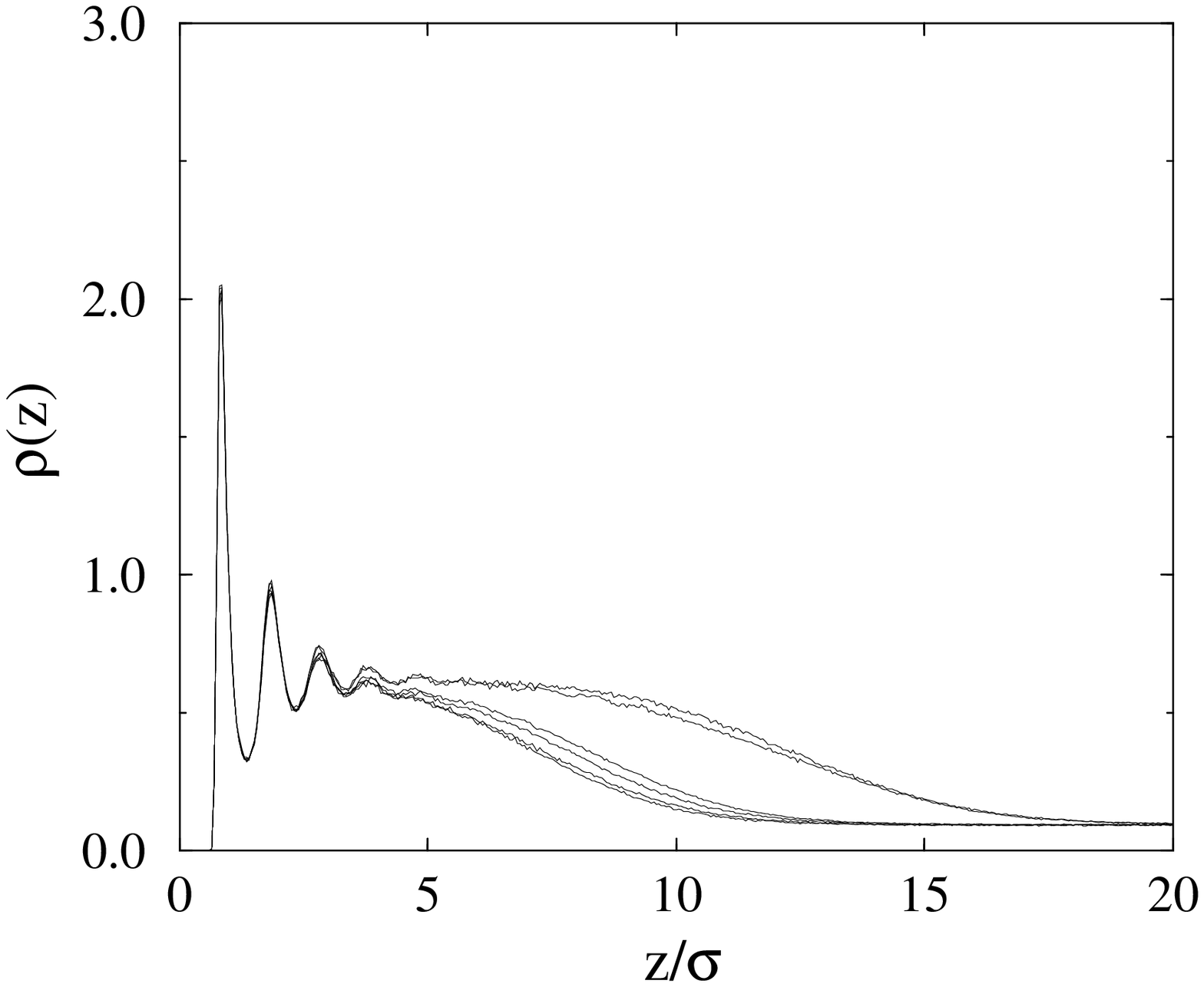}{65}{55}{
\vspace*{1.cm}
}

\noindent
\fig{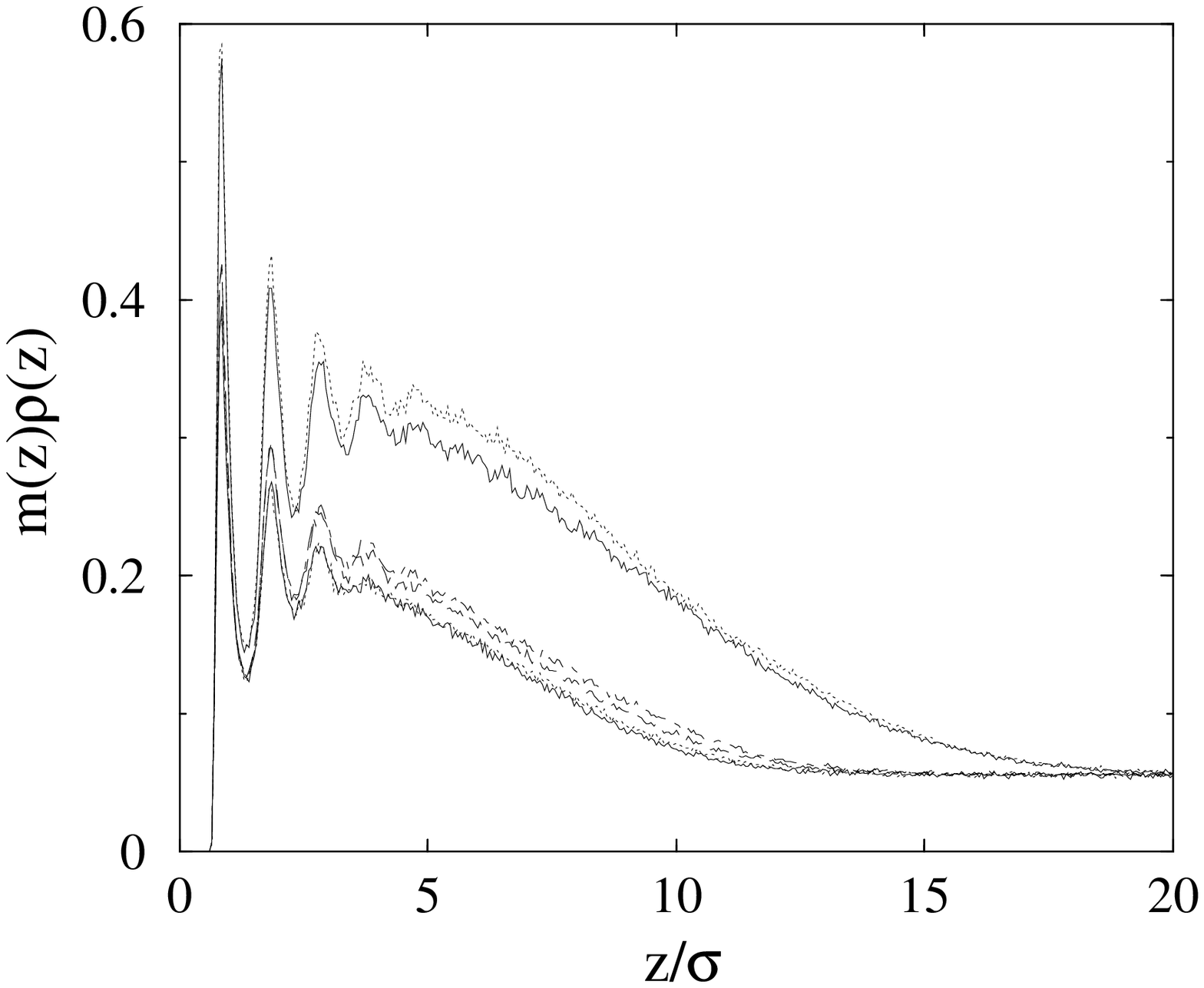}{65}{50}{
\vspace*{1.cm}
}
\CCewE
\end{figure}

\begin{figure}[t]
\noindent
\fig{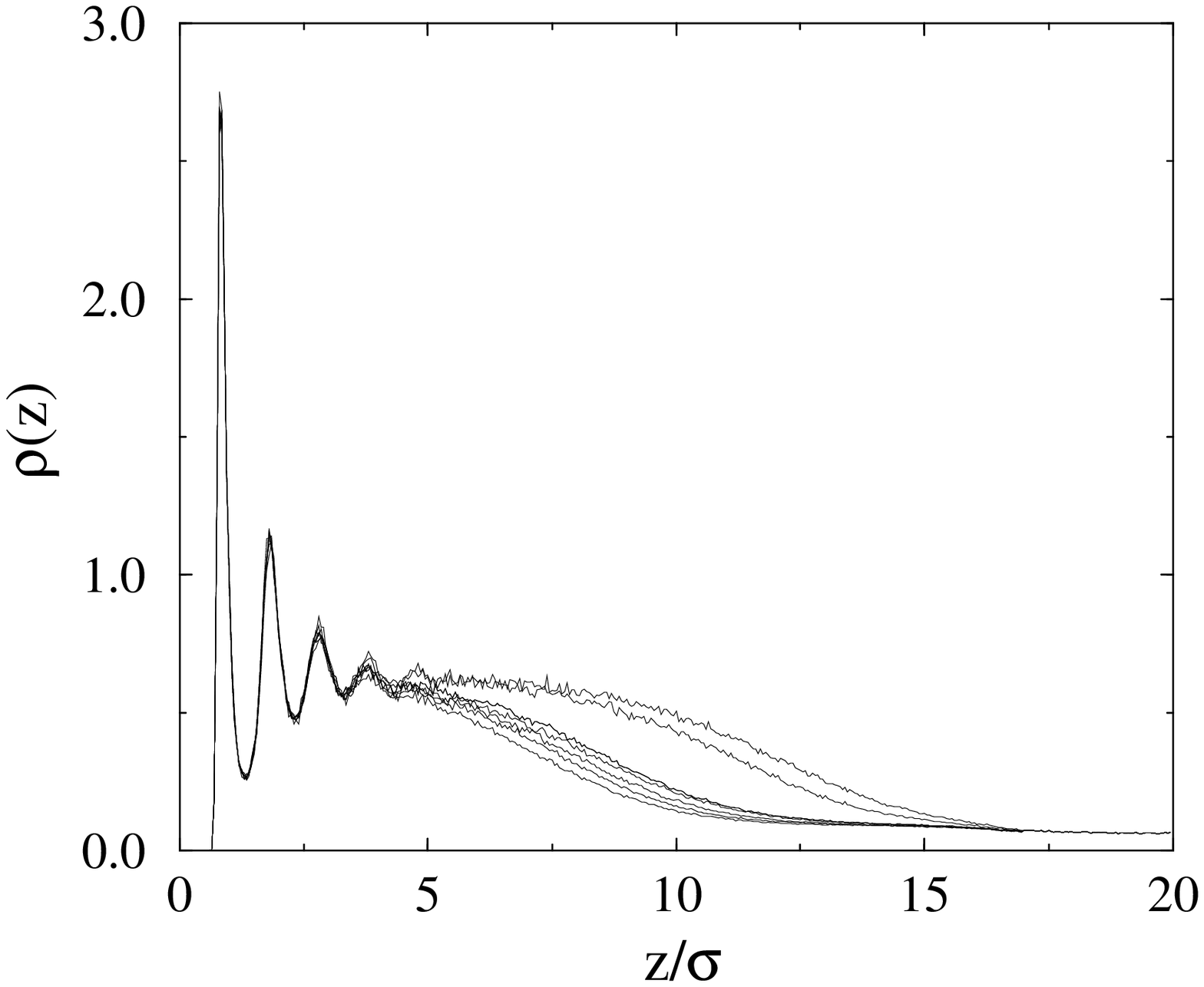}{65}{50}{
\vspace*{1.cm}
}
\CCewF
\end{figure}

\section{Discussion}
\label{sec:concl}

The Monte Carlo simulation results at subcritical temperatures provide 
evidence that the mean field calculations correctly identify the qualitative 
wetting behaviour. They show that depending on the
fluid-wall interaction strength $\epsilon_w$, a number of different
wetting scenarios occur  as liquid-vapour coexistence is approached from
the vapour side.  At small $\epsilon_w$, only a very thin film builds up
on the wall. For intermediate values of $\epsilon_w$, a first prewetting
transition is observed from a thin mixed film to a thick liquidlike
mixed layer. Further increasing $\epsilon_w$ induces a second prewetting
transition between a mixed liquidlike layer and a thicker demixed film,
the situation being very similar to that shown in
figure~\ref{fig:profiles}. The abrupt, first order, character of this
latter transition appears to weaken on further increasing $\epsilon_w$,
in accord with the theoretical predictions.

We will now attempt to set our results within the context of
the bulk phase diagram of the binary liquid. To this end, we discuss the
possible wetting scenarios in the vicinity of the critical end point
$T_{cep}$.  As previously argued in the introduction, for temperatures
$T < T_{cep}$ sufficiently close to $T_{cep}$, the bulk correlation
length $\xi$ of the demixed liquid is larger than the thickness $l_*$ of
a mixed liquid layer at the wall. The state of order of the film thus
depends strongly on the boundary conditions of the two interfaces
confining the liquid layer. The nonselective liquid-vapor interface
always favors mixing due to the reduced number of interacting neighbors
in the interfacial region. The liquid-substrate interface, on the other
hand, can either favour mixing or demixing depending on the strength of
the fluid-wall potential. For a weakly attractive wall potential, mixing
is favoured because the particle density at the wall is low and the
presence of the wall reduces the number of interacting neighbours. For a
strongly attractive wall, however, the high density at the wall can 
counteract the missing neighbour effect leading to an overall demixing
tendency.

\noindent
\fig{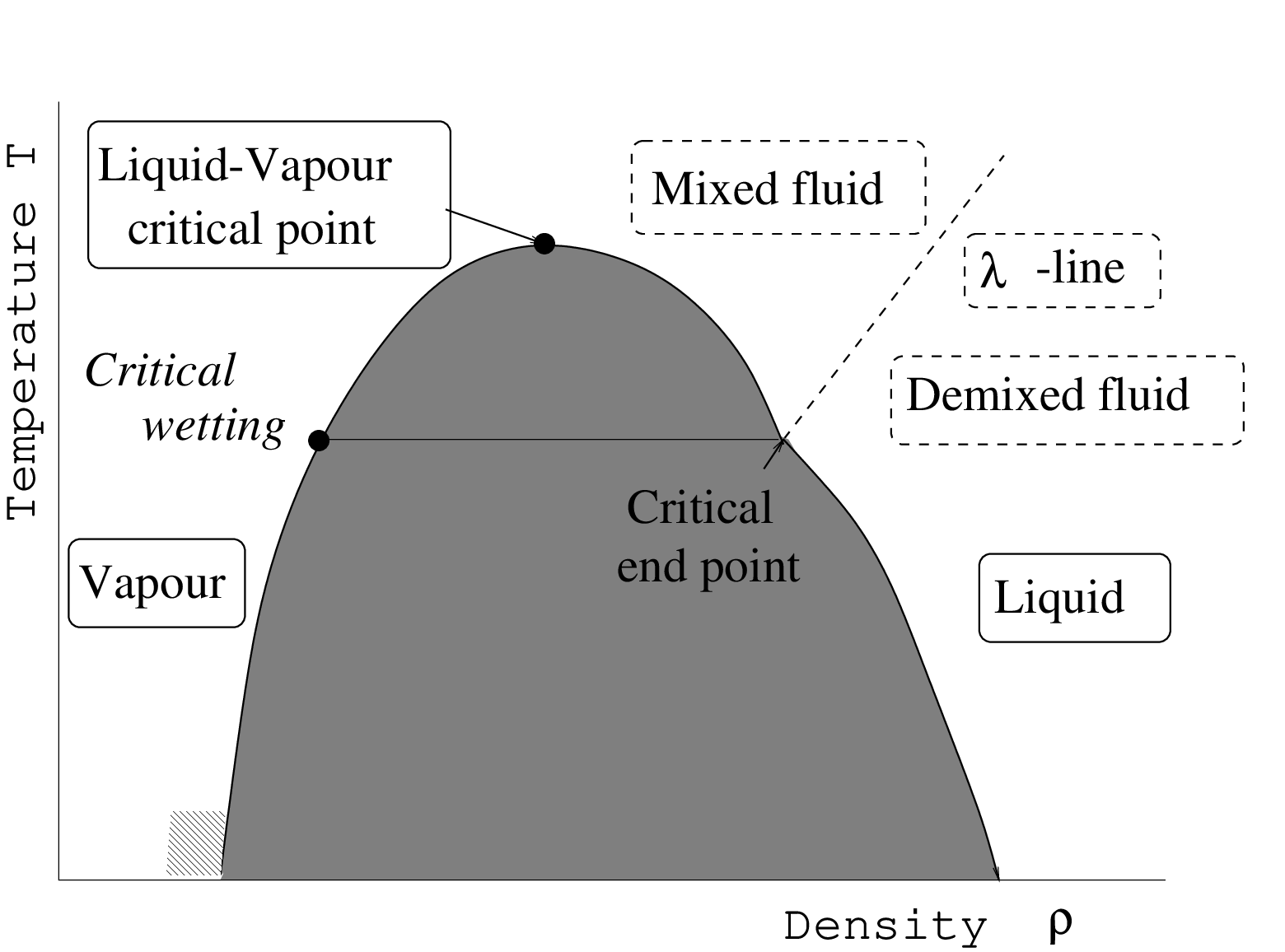}{75}{55}{
}

\noindent
\fig{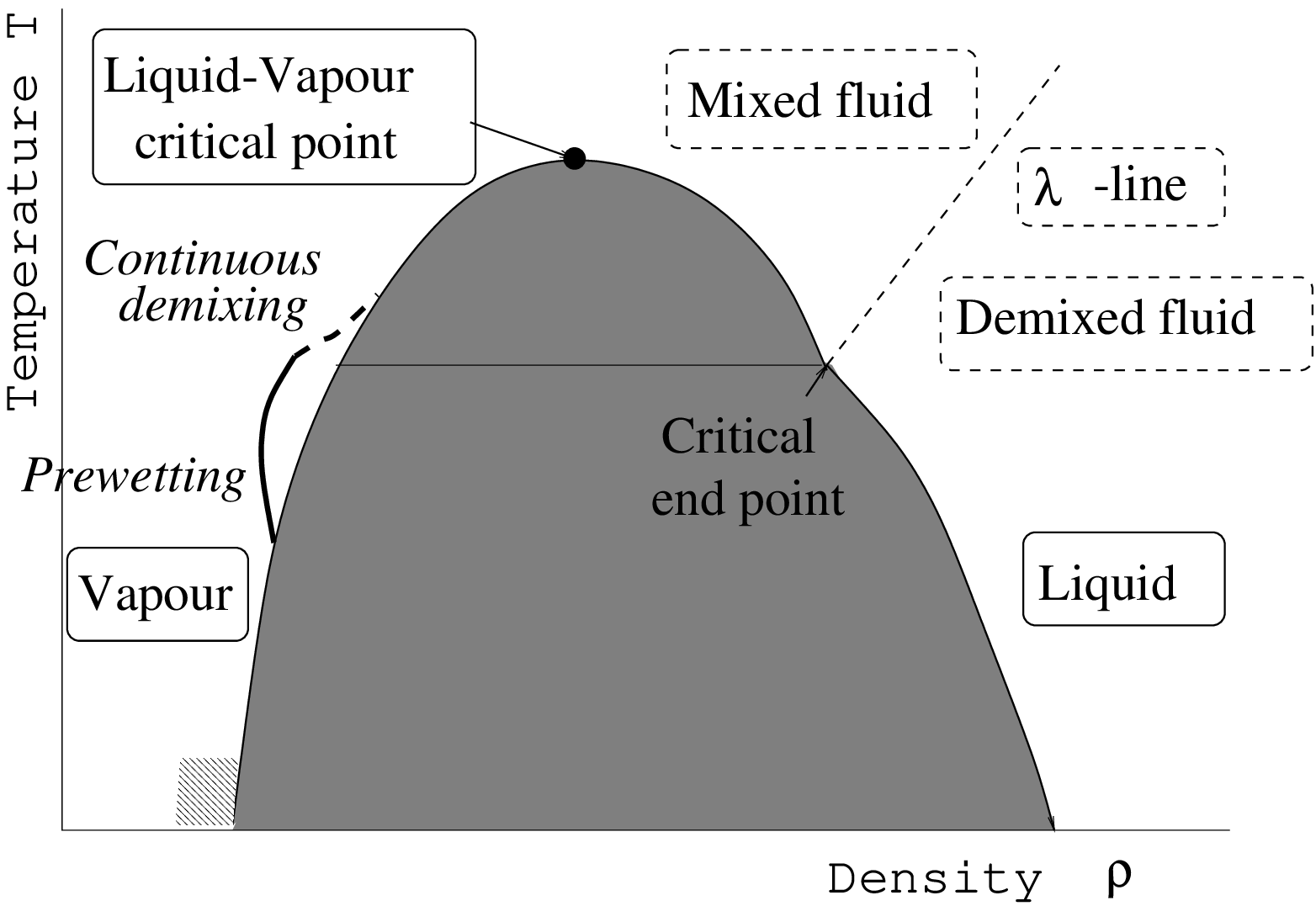}{75}{60}{
}

\begin{figure}[t]

\noindent
\fig{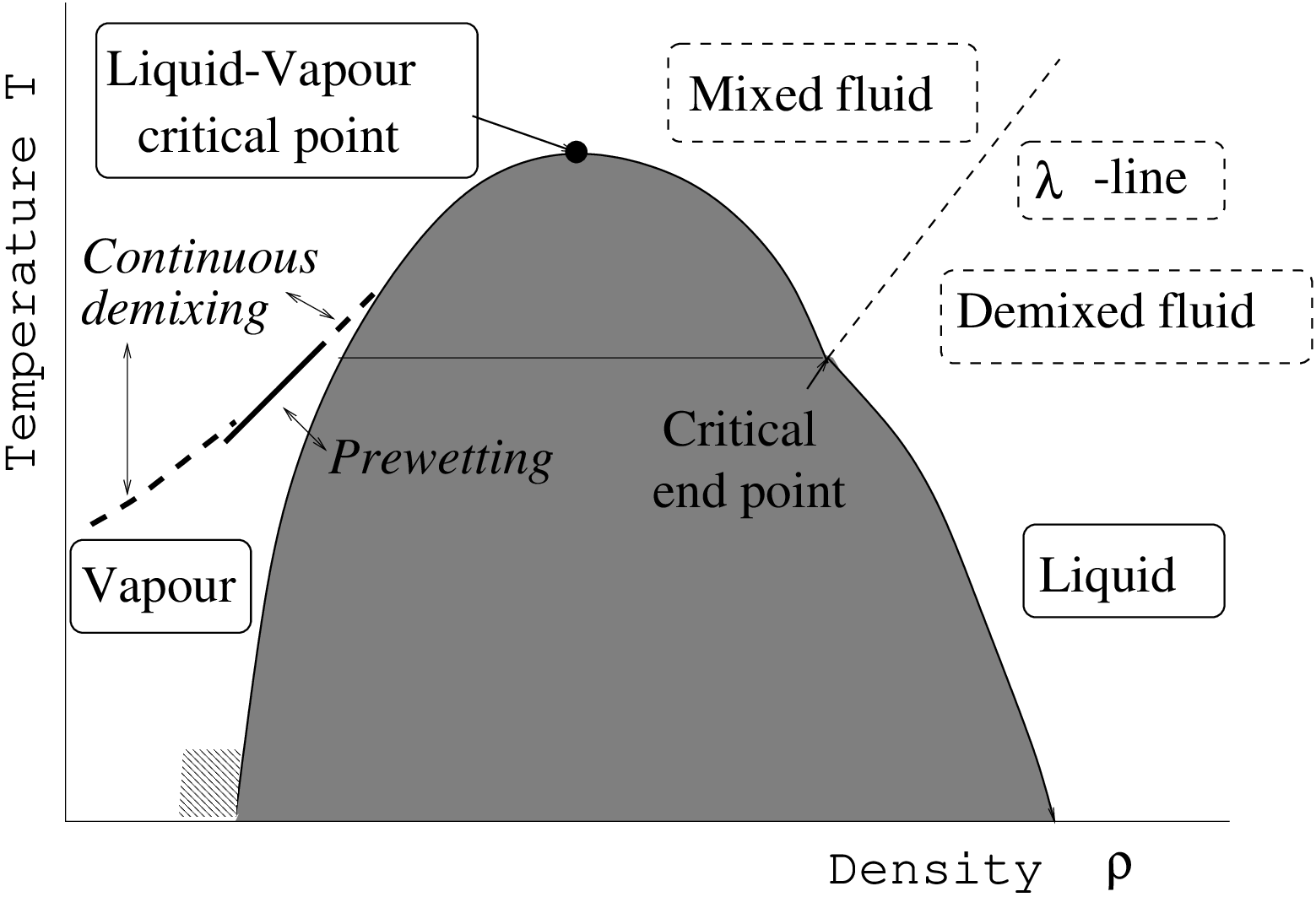}{75}{60}{
\vspace*{1.cm}
}
\CCwetschem
\end{figure}

If the net effect favours mixing at the wall, a continuous demixing of
the layer as coexistence is approached can be excluded. A first order
transition involving a discontinuous increase of the film thickness
upon demixing is still conceivable. However, we have shown in section
\ref{sec:limit}, that (at the mean field level, at least) the
demixed wetting film has a higher free energy than the corresponding
mixed film provided the correlation length of composition fluctuations
is sufficiently large.

At walls which suppress demixing, the film is thus always mixed close
to the critical end point, and its thickness $l_*$ below the critical
end point is finite. Hence the critical end point is automatically a
critical wetting point. The resulting phase diagram is shown
schematically in fig. \ref{fig:wetschem} (a). Note that the wetting
transition here is pinned by a bulk phase transition, a situation
somewhat reminiscent of triple-point wetting \cite{fisher2,dietrich3}.

The situation changes if the substrate favours demixing. In this
situation, one component segregates to the surface of the film already
slightly above $T_{cep}$, and the order propagates continuously into
the bulk of the film at $T_{cep}$.  The film remains wet at $T_{cep}$.
From the results of section \ref{sec:num} (in particular, Fig.
\ref{fig:phwet3}), one can deduce two possible scenarios. The film may
still exhibit a first order wetting transition to a nonwet state at a
temperature below $T_{cep}$  (e.g. in Fig. \ref{fig:phwet3}) at
$\phi_0 = 1.14$). The discontinuous phase transition at liquid/vapour
coexistence then spawns a prewetting line  which eventually switches
into a second order demixing line and loops around  the critical end
point as suggested in Fig. \ref{fig:wetschem} (b).  If the wall is
strongly attractive  (e.g. at $\phi_0 = 1.27$) in Fig.
\ref{fig:phwet3}), the wall wets at all temperatures, the prewetting
line detaches from the coexistence  line and is continued by second
order demixing lines both at the high  and low temperature side as
sketched in Fig. \ref{fig:wetschem} (c). 

\subsection*{Acknowledgements}

NBW thanks the Royal Society of Edinburgh, the EPSRC (grant no. GR/L91412) and
the British Council for financial support.


\begin{thebibliography}{50}
\bibitem{footnote}
 Quantitative difference, e.g., in the prefactors, are due to the
 fact that the underlying potentials $f(m,\phi)$ differ from each
 other.
\end{thebibliography}
\end{document}